\documentclass[12pt,a4paper]{article}
\usepackage[utf8]{inputenc}  %% add \usepackage[T1]{fontenc} if your TeX has the EC/LM Type1 fonts
\usepackage{amsmath,amssymb}
\usepackage{mathptmx}
\usepackage[super,numbers,sort&compress]{natbib}
\usepackage{booktabs}
\usepackage{longtable}
\usepackage{graphicx}
\usepackage[margin=2.5cm]{geometry}
\usepackage{setspace}
\usepackage{float}
\usepackage{caption}
\usepackage{authblk}
\usepackage{xcolor}
\usepackage[hidelinks]{hyperref}

\newcommand{\tabnote}[1]{\par\vspace{3pt}\noindent\begin{minipage}{\textwidth}\begin{spacing}{1}\footnotesize\raggedright\textit{Note:} #1\end{spacing}\end{minipage}}

\title{\textbf{BOP2-ENR: a Bayesian optimal phase II design for adaptive enrichment with family-wise error rate control and a bound on joint-claim power}}
\author[1]{Kentaro Takeda\textsuperscript{*}}
\author[2]{Belay B. Yimer}
\author[3]{Masahiro Kojima}
\affil[1]{Quantitative Sciences and Evidence Generation, Astellas Pharma Global Development Inc., Northbrook, IL, USA}
\affil[2]{Quantitative Sciences and Evidence Generation, Astellas Pharma Europe Ltd., Addlestone, UK}
\affil[3]{Department of Data Science for Business Innovation, Chuo University, Bunkyo-ku, Tokyo, Japan}
\affil[ ]{\normalfont *: Corresponding author: \href{mailto:kentaro.takeda@astellas.com}{kentaro.takeda@astellas.com}}
\date{}
\newcommand{\SIsec}[1]{\phantomsection\section*{#1}\addcontentsline{toc}{section}{#1}}
\newcommand{\SImark}[1]{\phantomsection}
\begin{document}
\addtocontents{toc}{\protect\setcounter{tocdepth}{-2}}
\maketitle

\doublespacing

\begin{abstract}
The Bayesian optimal phase II (BOP2) design accommodates simple and complex endpoints but does not address biomarker-guided enrichment or multiplicity control across pooled and subgroup claims. We propose BOP2-ENR, which monitors biomarker-negative (NG) and biomarker-positive (PG) cohorts separately, permits interim enrichment to PG, and calibrates pooled and PG-restricted claims. Strong family-wise error rate (FWER) control is defined over a prespecified monotone-activity parameter space. Under the mixed configuration in which NG is inactive and PG is active, the probability of a false joint claim is bounded by the probability that NG survives futility monitoring. This exact single-cohort bound does not involve the final efficacy cutoff and also yields a ceiling on joint-claim power. For efficacy-toxicity monitoring, subgroup inactivity is defined by a union null and safety is established separately within each subgroup. Across the scenarios considered and binary, co-primary and efficacy-toxicity endpoints, all selected designs met the prespecified exact-bound and Monte Carlo verification criteria; in sensitivity analyses over odds ratios between component outcomes assumed independent in calibration, point estimates of the error-control quantities remained below the nominal level, although power was association-sensitive. The bound was nearly attained as PG efficacy approached the boundary of the nuisance space, whereas disabling enrichment produced mixed-configuration error rates near one. BOP2-ENR therefore extends the BOP2 framework to adaptive enrichment while clarifying that the NG futility gate governs the mixed-configuration component of strong error control and the attainable power of a joint claim.
\end{abstract}

\noindent\textbf{KEYWORDS:} adaptive enrichment, Bayesian optimal phase II design, biomarker-stratified trial, family-wise error rate, nuisance parameter profiling, precision oncology
\section{Introduction}\label{sec1}

The Bayesian optimal phase II (BOP2) design \citep{Zhou2017-tv} has become an attractive framework for single-arm oncology trials. It models the outcome as multinomial with a Dirichlet prior, bases all go/no-go decisions on posterior probabilities of linear combinations of the model parameters, allows the probability cutoffs to depend on the accumulated information fraction, and controls the type I error rate explicitly while maximizing power. A single formalism thereby covers a binary response endpoint, nested endpoints, co-primary endpoints, and joint monitoring of efficacy and toxicity. Subsequent extensions have added time-to-event endpoints \citep{Zhou2020-gh,Lin2020-wz}, randomized comparisons \citep{Zhao2022-xr}, dual-criterion decision making \citep{Zhao2023-hu}, and efficacy stopping \citep{Xu2025-xq}.

What the BOP2 family does not address is population heterogeneity. The subgroup-unstratified BOP2 formulations considered here treat the trial population as homogeneous, with a single parameter vector $\theta$ and a single claim. In precision oncology, this is often untenable. When a predictive biomarker partitions patients into biomarker-positive (PG) and biomarker-negative (NG) subgroups and benefit is expected to concentrate in PG, a homogeneous design either dilutes a real effect by pooling indiscriminately or forgoes the broader claim by enriching from the outset. Neither is satisfactory when both an overall claim and a subgroup claim are scientifically plausible and the choice between them should be data-driven.

Adaptive enrichment addresses precisely this, and a well-developed line of work exists within the frequentist Simon two-stage tradition \citep{Simon1989-wl}. Jones et al.\citep{Jones2007-vv} adapted Simon's design to two parallel subgroups with interim routing between enrichment and unselected continuation, and Tournoux-Facon et al.\citep{Tournoux-Facon2011-zj} proposed a stratified design with an explicit heterogeneity check. Parashar et al.\citep{Parashar2016-de} formalized power definitions and introduced weak and strong control of the family-wise error rate \citep{Proschan2020-wr}. Most recently, the BOOST design \citep{Tong2026-es} integrated Simon-style Stage 1 futility rules, adaptive routing, a unified final-stage decision algorithm covering pooled and subgroup-specific claims, and exact FWER calibration within a constrained optimization framework. These designs, however, inherit the constraints of the Simon tradition: a binary endpoint, a single interim analysis, and count-scale decision rules. They cannot accommodate the co-primary or efficacy-toxicity structures that motivate BOP2, nor the flexible interim schedules that BOP2 permits.

The two literatures are therefore complementary, and within the peer-reviewed literature the gap is one-directional. The Simon-derived designs have the multiplicity framework but not the endpoint flexibility; BOP2 has the endpoint flexibility but not the multiplicity framework. A complementary direction is developed here: adaptive enrichment and pooled/subgroup FWER control are brought into BOP2, so that the resulting design retains BOP2's endpoint generality, interim flexibility, and posterior-scale decision rules while supporting biomarker-stratified claims with calibrated error control.

The result is called BOP2-ENR (a Bayesian optimal phase II design for adaptive enrichment). Its contributions are threefold. First, the BOP2 probability model is extended to two subgroups and adaptive enrichment is embedded in the interim monitoring, so that the trial can stop entirely, freeze the NG cohort and enrich to PG, or continue both to a pooled analysis. Second, weak and strong FWER are defined and calibrated across the pooled and subgroup claims, using a computational identity that keeps complex endpoints tractable. Third, the sources of strong control in designs of this class are distinguished: the NG futility gate controls the mixed-configuration component and the final efficacy cutoff the global-null component. The mixed-configuration error rate is shown to be bounded by the probability that the NG cohort survives futility monitoring, a bound that does not involve the final efficacy boundary; in the settings examined the bound was approximately attained at the boundary of the nuisance parameter space rather than at the design alternative, and the same gate is shown to cap the power of the joint claim. A design calibrated without this observation can satisfy weak control while violating strong control by nearly fivefold.

{The remainder of the paper is organized as follows. Section~\ref{sec2} develops the design: the probability model, the endpoints, the enrichment routing, the error rates, the bound on the mixed-configuration error rate, and the sequential calibration that it enables. Section~\ref{sec3} reports the simulation study, covering the common settings and comparators, the behavior of the bound along the nuisance path, and the operating characteristics for single binary, co-primary and efficacy-toxicity endpoints. Section~\ref{sec5} discusses implications and limitations.}

\section{Methods}\label{sec2}

\subsection{Probability model}\label{sec2.1}

Following Zhou et al.\citep{Zhou2017-tv}, the outcome of each patient is assumed to be a random variable $Y$ taking values in $K$ categories,

\begin{equation*}
Y \sim \mathrm{Multinomial}(\theta_1, \dots, \theta_K), \qquad \theta \sim \mathrm{Dir}(a_1, \dots, a_K),
\end{equation*}

where $\theta_k = \Pr(Y = k)$ is the probability that $Y$ falls in the $k$th category, $k = 1, \dots, K$, and $\sum_k a_k = 1$ gives a vague prior carrying the weight of a single observation. Following the BOP2 convention the prior is centered on the null: the concentration parameters are chosen so that each monitored quantity has prior mean equal to its own null threshold, $\sum_{k \in \mathbf{b}_j} a_k = \phi_{0j}$ for every criterion $j$, where $\mathbf{b}_j \in \{0,1\}^K$ is the design vector selecting the categories aggregated by criterion $j$ (the criteria for each endpoint are listed below). Each criterion $j$ thus compares a sum of category probabilities, $\mathbf{b}_j\theta$, with its null threshold $\phi_{0j}$: for the single binary endpoint $K = 2$ and the only criterion is $\theta_1 > \phi_0$ with $\mathbf{b}_1 = (1,0)$, whereas for the co-primary endpoints the response criterion compares $\theta_1 + \theta_2$ with $\phi_{01}$, so that $\mathbf{b}_1 = (1,1,0,0)$. Since the criteria are subset sums rather than individual categories, this pins down the prior mass of each monitored set rather than the individual $a_k$; for the binary endpoint it gives $\mathrm{Beta}(\phi_0, 1 - \phi_0)$, and for the two-criterion endpoints any $a$ satisfying the two constraints yields identical decisions, because every rule depends on $a$ only through these subset sums. The posterior after observing counts $D_n = (x_1, \dots, x_K)$ is $\theta \mid D_n \sim \mathrm{Dir}(a_1 + x_1, \dots, a_K + x_K)$. The beta-binomial model, that arises from binary-endpoint designs, is the special case with $K = 2$. Throughout, we use the general Dirichlet-multinomial formulation, with the binary endpoint setting treated as one instance of the general framework.

BOP2-ENR stratifies this model by biomarker status, positing separate parameter vectors $\theta^-$ and $\theta^+$ for NG and PG with independent Dirichlet priors, and maximum sample sizes $N^-$ and $N^+$. Following the convention of adaptive enrichment designs in this setting, a monotone efficacy ordering is assumed, and it is a restriction on the parameter space rather than an expectation: for every efficacy criterion, $\mathbf{b}\theta^- \le \mathbf{b}\theta^+$, that is, PG is at least as responsive as NG. Together with the monotone-activity assumption introduced in Section~\ref{sec2.4} it defines the admissible space over which strong control is claimed; no ordering is assumed for a toxicity criterion. The two restrictions act at different levels: the monotone-activity assumption is stated in terms of activity, which for efficacy-toxicity monitoring requires a subgroup to be both efficacious and safe, and therefore excludes configurations in which NG is active (efficacious and safe) while PG is inactive on either face; it imposes no ordering of the toxicity rates themselves between subgroups.

Decisions are based on posterior probabilities of linear combinations $\mathbf{b}\theta$, with $\mathbf{b}$ one of the design vectors $\mathbf{b}_j$ above (the subscript is dropped when a single criterion is in view). The computational device on which the whole construction rests is that a subset sum of a Dirichlet vector is Beta distributed:

\begin{equation*}
\mathbf{b}\theta \mid D_n \sim \mathrm{Beta}\!\left(\sum_{k \in \mathbf{b}} (a_k + x_k),\;\; \sum_{k \notin \mathbf{b}} (a_k + x_k)\right).
\end{equation*}

Consequently, $Pr(\mathbf{b}\theta > \phi_0 \mid D_n)$ is available in closed form as a Beta tail probability of aggregated counts, with no numerical integration over $\theta$ and no posterior sampling. Since the aggregated counts are discrete, the required tail probabilities can be tabulated once per look rather than evaluated for every simulated trial, which is what makes calibration over multiple endpoints and configurations feasible.

\subsection{Endpoints}\label{sec2.2}

Three of the endpoint structures of the original BOP2 design are considered \citep{Zhou2017-tv}. Null thresholds may differ between subgroups and then carry the superscript $\pm$ (Section~\ref{sec2.3}); the toxicity threshold is common to both subgroups.

\textbf{Single binary endpoint} ($K = 2$; $1$ = response, $2$ = no response). One criterion, $\theta_1 > \phi_0$.

\textbf{Co-primary endpoints} ($K = 4$; categories formed by crossing response with six-month progression-free survival, PFS6). Two criteria, $\theta_1 + \theta_2 > \phi_{01}$ and $\theta_1 + \theta_3 > \phi_{02}$. The treatment is futile only if \textit{both} criteria fail; promising if \textit{either} succeeds. The term co-primary is inherited from BOP2 \citep{Zhou2017-tv} and denotes this disjunctive success rule, in which either endpoint suffices for a claim; it differs from the regulatory usage in which both endpoints must succeed.

\textbf{Joint efficacy-toxicity monitoring} ($K = 4$; categories formed by crossing response with toxicity). Two criteria, $\theta_1 + \theta_2 > \phi_{01}$ (efficacy) and $\theta_1 + \theta_3 < \phi_{02}$ (toxicity). The treatment is futile if \textit{either} criterion fails; promising only if \textit{both} succeed. The toxicity criterion is directional in the opposite sense, which the Beta representation handles by using the distribution function. Because a claim requires both criteria, a subgroup is inactive for this endpoint whenever either fails: with the superscripts $-$ and $+$ indexing NG and PG throughout, its null set is the union $H_0^{\pm} = \{\theta_E^{\pm} \le \phi_{01}^{\pm}\} \cup \{\theta_T^{\pm} \ge \phi_{02}\}$, and strong control must hold on both faces of this union, the efficacy-null face (inefficacious, however safe) and the toxicity-null face (unacceptably toxic, however efficacious). Two consequences for the decision rules follow: safety is established within each subgroup and is never pooled across subgroups, and the toxicity criterion carries its own claim cutoff $\lambda_T$, calibrated exactly against the toxicity null of the cohort size concerned, in place of the efficacy cutoff $\lambda_E$. For both multi-criterion endpoints the two component outcomes are modeled as independent given their marginals; this is a design assumption under which the designs are calibrated, and its role and the sensitivity to association are discussed in Section~\ref{sec3.1}.

The futility and efficacy combination operators are written generically as $\mathrm{op}_F$ and $\mathrm{op}_E$, taking the values (AND, OR) for binary and co-primary endpoints and (OR, AND) for efficacy-toxicity.

\subsection{Trial conduct and enrichment routing}\label{sec2.3}

Both cohorts accrue concurrently on a shared timeline indexed by information fractions $t_1 < \cdots < t_R = 1$, so that at look $r$ the enrolled sizes are $\lceil t_r N^+ \rceil$ and $\lceil t_r N^- \rceil$. Concurrent accrual must be modelled explicitly, since a futility signal in PG halts enrollment in both cohorts and the expected sample size depends on this coupling.

At each interim look, subgroup-specific futility boundaries are applied:

\begin{equation*}
C_F^-(n) = \lambda_- \left(\frac{n}{N^-}\right)^{\gamma_-}, \qquad C_F^+(n) = \lambda_+ \left(\frac{n}{N^+}\right)^{\gamma_+}.
\end{equation*}

Allowing the two subgroups to have distinct tuning parameters is essential rather than cosmetic: $(\lambda_-, \gamma_-)$ is the lever that bounds the mixed-configuration error rate, while $(\lambda_+, \gamma_+)$ governs efficiency.

The routing algorithm is as follows.

\begin{itemize}
\item \textbf{Stop.} If the PG criteria collectively signal futility (via $\mathrm{op}_F$), the trial terminates and no claim is made.
\item \textbf{Enrich.} If the NG criteria signal futility while PG continues, NG accrual is closed at its current size and PG accrues to $N^+$; after NG accrual is frozen, PG continues on its prespecified interim schedule and may still stop for futility before reaching $N^+$. The NG cohort is described throughout as \textit{frozen} rather than dropped: no further NG patients are enrolled; those already enrolled remain under follow-up, but no pooled claim is evaluated after enrichment. At the final look a PG-restricted claim is evaluated.
\item \textbf{Continue both.} Otherwise both cohorts accrue to their maxima and a pooled analysis is performed at the final look, with a PG-restricted fallback if the pooled claim fails.
\end{itemize}

The claims available on each path are therefore: none after a stop; the PG-restricted claim, and no pooled claim, after enrichment; and the pooled claim with PG-restricted fallback after continue-both. The pooled claim tests the allocation-weighted mixture estimand $w\,\mathbf{b}\theta^- + (1-w)\,\mathbf{b}\theta^+$ against its mixture threshold, with $w = N^-/(N^- + N^+)$. Asserting benefit for NG on the basis of that claim is a decision-level joint claim: it combines the pooled test with the gatekeeping supplied by the NG futility monitoring that the continue-both path presupposes, and it is not an independent posterior demonstration of efficacy in NG. The error-rate definitions and Propositions (a) and (b) below are stated for this decision rule; a variant requiring additional NG-specific evidence alongside the pooled claim would inherit the same bound, since it can only remove joint claims. For joint efficacy-toxicity monitoring this variant is the design: the pooled statistic is used for the efficacy criterion only, and the joint claim additionally requires the subgroup-specific safety criteria $\Pr(\theta_T^- < \phi_{02} \mid D^-) \ge \lambda_T^-$ and $\Pr(\theta_T^+ < \phi_{02} \mid D^+) \ge \lambda_T^+$ at the final analysis, while the PG-restricted claim requires $\Pr(\theta_T^+ < \phi_{02} \mid D^+) \ge \lambda_T^+$ together with its efficacy criterion. A pooled safety statistic is not used because it tests a mixture of the two toxicity rates, so that a safe cohort could carry an unacceptably toxic one to a claim in either direction. A route in which NG alone continues is excluded by the monotone-activity assumption. A structural consequence used repeatedly is that an NG claim is never made alone: NG efficacy is asserted only jointly with PG, through a significant pooled analysis.

Final claims use a common posterior cutoff $\lambda_E$ for the efficacy criteria. For the single binary and co-primary endpoints, this is the only final cutoff. For joint efficacy-toxicity monitoring, $\lambda_E$ governs only the efficacy criterion, and the subgroup-specific safety cutoffs $\lambda_T^{\pm}$ described above are applied in addition. The PG-restricted claim is made if the PG efficacy criteria satisfy $\mathrm{op}_E$ at cutoff $\lambda_E$; the pooled claim is made if the efficacy criteria evaluated using the combined counts do so. For asymmetric settings in which the subgroup null thresholds differ, the pooled null threshold is taken as the allocation-weighted mixture $\phi_0^{\mathrm{pool}} = w\phi_0^- + (1-w)\phi_0^+$ with $w = N^-/(N^- + N^+)$, and the pooled prior is formed correspondingly. The pooled Beta tail is a decision statistic based on a working pooled model; it is not the posterior distribution of the allocation-weighted mixture induced by the two independent subgroup posteriors, and all operating characteristics are evaluated under subgroup-specific data-generating distributions.

\subsection{Error rates and power}\label{sec2.4}

Under weak control the FWER is the probability of at least one efficacy claim under the global null, and disjunctive power is the probability of at least one claim under the global alternative.

Under strong control the error rate must be bounded across all admissible configurations of true and false subgroup nulls. The admissible space is defined by a monotone-activity assumption stated at the level of the hypotheses themselves: if NG is active then PG is active. This is the biological premise of enrichment, that the biomarker is predictive and benefit is concentrated in PG, and it is stronger than an ordering of raw response rates: in asymmetric settings ($\phi_0^- \ne \phi_0^+$) a common rate $p^- = p^+ \in (\phi_0^-, \phi_0^+]$ satisfies $p^+ \ge p^-$ yet has NG active and PG null, a configuration the assumption excludes. Writing $\Theta_{00}, \Theta_{01}, \Theta_{10}$ for the sets with both subgroups null, NG null with PG active, and NG active with PG null, the strong-control error rate is $\max\{\sup_{\Theta_{00}} P(\text{any false claim}), \sup_{\Theta_{01}} P(\text{false NG claim}), \sup_{\Theta_{10}} P(\text{false PG claim})\}$; under the assumption $\Theta_{10}$ is excluded from the admissible parameter space, and it suffices to consider the global null and the mixed configuration in which NG is inactive and PG is active. Throughout, "strong FWER control" means control over this prespecified monotone-activity parameter space, not over the unrestricted space:

\begin{equation*}
\mathrm{FWER}_{00} = \sup_{(\theta^-, \theta^+) \in \Theta_{00}} P(\text{claim NG or PG} \mid \theta^-, \theta^+),
\end{equation*}

\begin{equation*}
\mathrm{FWER}_{01}(\psi) = P(\text{claim NG and PG} \mid \theta^- = \theta_0^-, \theta^+ = \psi), \quad \psi \in \Theta_1^+,
\end{equation*}

\begin{equation*}
\mathrm{FWER}_S = \max\{\mathrm{FWER}_{00},\; \textstyle\max_{\mathrm{grid}}\mathrm{FWER}_{01}\},\qquad U_S = \max\{\mathrm{FWER}_{00},\; P(S)\},
\end{equation*}

The subscripts index the configuration: $00$ for the global null $\Theta_{00}$, $01$ for the mixed configuration $\Theta_{01}$ (NG null, PG active) and $11$ for the global alternative $\Theta_{11}$. The conjunctive form of $\mathrm{FWER}_{01}$ follows from the claim structure: since NG is claimed only jointly with PG, any false NG claim is necessarily a joint claim. Table~\ref{tab0} summarizes the configurations, the claims available on each path (Section~\ref{sec2.3}), and which of them are erroneous in each configuration.

The nuisance parameter $\psi$ is not a design input; it is profiled out by grid search along a path interpolating from the null vector to an extreme alternative in which the favorable categories carry nearly all mass. For efficacy-toxicity monitoring, where the subgroup null is a union, $\theta_0^{\pm}$ in these definitions denotes the least-favorable point of each face as stated in the Lemma below. Specifically, $\mathrm{FWER}_{00}$ is the maximum of the any-claim probability over the admissible corners of $H_0^- \times H_0^+$, and $\mathrm{FWER}_{01}(\psi)$ is evaluated with the NG cohort on either face. On the toxicity face, the NG efficacy is set equal to the PG efficacy $\psi_E$ so that the profile respects the ordering; the exact bound for that face is computed at NG efficacy one, the supremum over NG efficacy, and is therefore conservative for the admissible profile (numerically identical in every scenario). The four face combinations of $H_0^- \times H_0^+$ are all part of $\Theta_{00}$, since a cohort on its toxicity face is inactive however efficacious it is. Three of them are unrestricted, and their least-favorable points are the corner pairs of the face points above. The fourth, NG on its toxicity face and PG on its efficacy face, is cut by the efficacy ordering of Section~\ref{sec2.1}, $\theta_E^- \le \theta_E^+ = \phi_{01}^+$: within it the least-favorable point is NG at $(\phi_{01}^+, \phi_{02})$ and PG at $(\phi_{01}^+, 0)$, by the monotonicity of the Lemma. Without the ordering this combination would admit NG at $(1, \phi_{02})$, and the design would not be guaranteed to control the error rate there; the ordering is therefore a substantive part of the guarantee. Reported $\max_{\mathrm{grid}}\mathrm{FWER}_{01}$ values are maxima over this prespecified path grid, not certified suprema over the full admissible space. $\mathrm{FWER}_S$ is therefore a numerical diagnostic of the strong-control error rate, not the guaranteed quantity. The guarantee is carried by $U_S$, where $P(S)$ is the upper bound on $\mathrm{FWER}_{01}(\psi)$ established in Section~\ref{sec2.5}, which holds uniformly over the admissible space and not only on the grid; the comparison tables report both $P(S)$ (the certifying bound) and $\max_{\mathrm{grid}}\mathrm{FWER}_{01}$ (the profiled diagnostic of its tightness). Because $\mathrm{FWER}_{00}$ is estimated by Monte Carlo whereas $P(S)$ is exact, $U_S$ is reported in two versions, a point estimate in the comparison tables and a verification version with multiplicity-adjusted upper confidence limits.

Power is defined configuration-specifically, with $\mathrm{POWER}_{01} = P(\text{claim PG} \mid \theta_0^-, \theta_1^+)$ under the mixed configuration and $\mathrm{POWER}_{11} = P(\text{claim NG and PG} \mid \theta_1^-, \theta_1^+)$ under the global alternative. Because NG is never claimed alone, $\mathrm{POWER}_{01}$ coincides with the probability of any claim under the mixed configuration. Two further summaries are reported: \textit{disjunctive power}, the probability of any claim under the global alternative (the weak-control power criterion), and $\mathrm{ESS}_0$, the expected sample size under the global null.

\subsection{A ceiling on the mixed-configuration error rate}\label{sec2.5}

\textbf{Lemma (least favorable point of the global null).} \textit{Suppose every criterion is coordinate-wise monotone: the posterior tail $T(x, n)$ is increasing in the aggregated count $x$ for a criterion with direction $>$ and decreasing for direction $<$. Then, for the single binary endpoint, $P(\text{any claim} \mid p^-, p^+)$ is non-decreasing in each of $p^-$ and $p^+$, so that $\sup_{\Theta_{00}} P(\text{any claim})$ is attained at the boundary point $(\theta_0^-, \theta_0^+)$.} The proof and the least-favorable points of each endpoint are given in Appendix S1; $\mathrm{FWER}_{00}$ is evaluated at those points throughout.

\textbf{Proposition (a) (upper bound on the mixed-configuration error rate).} \textit{Let $S$ denote the event that the NG cohort survives every interim futility look under $\theta^- = \theta_0^-$. Then, for every admissible $\psi$,}

\begin{equation*}
\mathrm{FWER}_{01}(\psi) \;\le\; P(S).
\end{equation*}

\textbf{Proof.} Let $C$ denote the event of an erroneous joint claim under $(\theta_0^-, \psi)$. A joint claim is available only on the continue-both path (Section~\ref{sec2.3}), which requires the NG cohort to survive every interim futility look; hence $C \subset S$ and $P(C) \le P(S)$. $\square$

\textbf{Interpretation.} A false joint claim can arise through one channel only: under the NG null, the NG cohort must first survive its own futility monitoring, and nothing downstream of that gate (the PG gate, the claim rules, the efficacy cutoff $\lambda_E$) can raise the joint-claim probability above $P(S)$. The bound is therefore a single-cohort quantity, computable exactly (Appendix S3), that does not involve $\lambda_E$ or $(\lambda_+, \gamma_+)$, and it holds uniformly in $\psi$ and hence for $\max_{\mathrm{grid}}\mathrm{FWER}_{01}$ whatever grid is used.

\textbf{Proposition (b) (the toxicity face).} \textit{For joint efficacy-toxicity monitoring, let $S_T$ denote the event that the NG cohort survives every interim look and satisfies its safety criterion $\Pr(\theta_T^- < \phi_{02} \mid D^-) \ge \lambda_T^-$ at the final analysis, under an NG cohort on its toxicity-null face. Then $\mathrm{FWER}_{01}(\psi) \le P(S_T) \le P(\text{NG safe} \mid \theta_T^- = \phi_{02})$, and the last quantity is an exact binomial tail that the calibration of $\lambda_T^-$ places at or below $\alpha$.} The proof is the same event inclusion: the joint claim requires the NG safety criterion, which is monotone in the toxicity count. On the efficacy-null face Proposition (a) applies unchanged with $P(S)$ evaluated at $(\phi_{01}, 0)$. The certifying bound for this endpoint is the larger of the two face values, and both are single-cohort quantities computed exactly by the recursion of Appendix S3.

The condition under which the bound is attained exactly, and its tightness in every scenario-by-endpoint combination, are given in Appendix S1.

\textbf{Corollary (a ceiling on joint power).} \textit{Let $S^{\ast}$ denote the event that the NG cohort survives every interim futility look under $\theta^- = \theta_1^-$. Then}
\begin{equation*}
\mathrm{POWER}_{11} \;\le\; P(S^{\ast}).
\end{equation*}
\textit{Proof.} Identical to that of Proposition (a): a joint claim is available only on the continue-both path, which requires the NG cohort to survive every interim look. $\square$

For efficacy-toxicity the joint claim also requires both subgroup safety criteria, so the ceiling tightens to $P(S^{\ast} \cap \text{NG safe}) \cdot P(\text{PG safe})$ under the tolerable toxicity rate, the two cohorts being independent. Proposition (a) and the Corollary are two faces of the same gate. The NG futility rule that bounds the mixed-configuration error rate by $P(S)$ under the NG null also bounds the joint power by $P(S^{\ast})$ under the NG alternative, and both bounds are single-cohort quantities that do not involve $\lambda_E$ or the PG tuning parameters. Calibrating $(\lambda_-, \gamma_-)$ so that $P(S) \le \alpha$ therefore fixes, at the same time, a ceiling on $\mathrm{POWER}_{11}$ equal to the power of that single-cohort futility rule under the NG alternative; the trade-off between the mixed-error bound and the joint-power ceiling is determined by the operating characteristics of the NG gate.

\subsection{Sequential calibration}\label{sec2.5b}

Proposition (a) separates the two components of the restricted strong FWER: the NG futility gate controls the mixed-configuration component through a bound that does not involve $\lambda_E$, so that tightening $\lambda_E$ cannot be relied on to control $\mathrm{FWER}_{01}$ (whenever the attainment condition of Appendix S1 holds, at the boundary of the nuisance path the pooled statistic is driven past the pooled-claim boundary by the PG cohort alone), whereas the final efficacy cutoff controls the global-null component. The bound is a single-cohort quantity computable without simulating the full trial, which makes it a direct calibration target.

Proposition (a) licenses a sequential calibration in place of a joint search over the tuning parameters:

\begin{itemize}
\item \textbf{Step 1 (error control).} Choose $(\lambda_-, \gamma_-)$ as the first feasible gate in a prespecified scan order over a grid, that is, the first pair whose exact bound $P(S)$ at the least-favorable NG null is at most $\alpha$; for efficacy-toxicity the efficacy-null face $(\phi_{01}, 0)$ is used. The bound is computed by the finite-state recursion of Appendix S3, so no Monte Carlo margin is needed.
\item \textbf{Step 1b (safety cutoffs, efficacy-toxicity only).} Choose $\lambda_T^-$ and $\lambda_T^+$ as the smallest values on a prespecified grid (Appendix S4) whose exact size at the toxicity null, $P(\Pr(\theta_T < \phi_{02} \mid x, N) \ge \lambda_T \mid \theta_T = \phi_{02})$, is at most $\alpha$ for $N = N^-$ and $N = N^+$ respectively. This bounds the toxicity face by Proposition (b).
\item \textbf{Step 2 (efficiency).} Choose $(\lambda_+, \gamma_+)$, which trades expected sample size against power without affecting the upper bound $P(S)$.
\item \textbf{Step 3 (calibration of the claim).} Choose the smallest $\lambda_E$ on a grid such that $\mathrm{FWER}_{00} \le 0.9\alpha$, the global-null error rate being estimated by Monte Carlo and, for efficacy-toxicity, maximized over the admissible null corners.
\end{itemize}

Among candidate tuning parameters meeting the power target ($\mathrm{POWER}_{01} \ge 0.80$ at the design alternative), the one minimizing the expected sample size under the global null is retained. ``Optimal'' is inherited from the BOP2 nomenclature; the present implementation is a sequential constrained grid selection with fixed sample sizes and does not claim joint global optimality over sample sizes and all tuning parameters. The margin $0.9\alpha$ applies only to the Monte Carlo constraint of Step 3; the exact constraints of Steps 1 and 1b are applied at $\alpha$ itself. The selected designs are then re-evaluated from an independent seed with a larger replicate count and one-sided upper confidence limits (Table~\ref{tabS12}).

\subsection{Prespecification of the decision boundaries}\label{sec2.6}

A practical requirement for a design intended for protocol use (and one of the properties that recommends BOP2) is that the stopping rules can be tabulated before the first patient is enrolled, so that the protocol states integer decision rules rather than a posterior computation to be performed at each interim.

BOP2-ENR retains this property, and the Beta representation of Section~\ref{sec2.1} makes the inversion immediate. For a criterion with direction $>$, the posterior tail $T(x, n) = Pr(\mathbf{b}\theta > \phi_0 \mid x \text{ of } n)$ is strictly increasing in the aggregated count $x$, so the rules invert to

\begin{equation*}
\text{futility } (T < C_F(n)) \iff x \le b_F(n), \qquad \text{claim } (T \ge \lambda_E) \iff x \ge b_E,
\end{equation*}

with $b_F(n) = \max\{x : T(x,n) < C_F(n)\}$ and $b_E = \min\{x : T(x,n) \ge \lambda_E\}$. For the toxicity criterion, whose direction is $<$, the tail is decreasing in $x$ and both inequalities reverse: futility corresponds to $x \ge b_F$ and a claim to $x \le b_E$. Multi-criterion endpoints are tabulated per criterion and combined by the endpoint's own operators, $\mathrm{op}_F$ and $\mathrm{op}_E$. For efficacy-toxicity the toxicity claim boundaries are tabulated with $\lambda_T^{\pm}$ in place of $\lambda_E$, separately for the NG cohort at $N^-$ and the PG cohort at $N^+$, and the joint claim requires both (Table~\ref{tab8}).

One feature is specific to the enrichment structure and deserves note. The pooled claim is evaluated only on the continue-both path (Section~\ref{sec2.3}), on which the NG cohort has accrued in full, so a single pooled boundary for the combined cohort of $N^- + N^+$ patients suffices and is prespecified alongside the others. Nothing remains to be computed during the trial.

\begin{figure}[htbp]
\centering
\includegraphics[width=\textwidth]{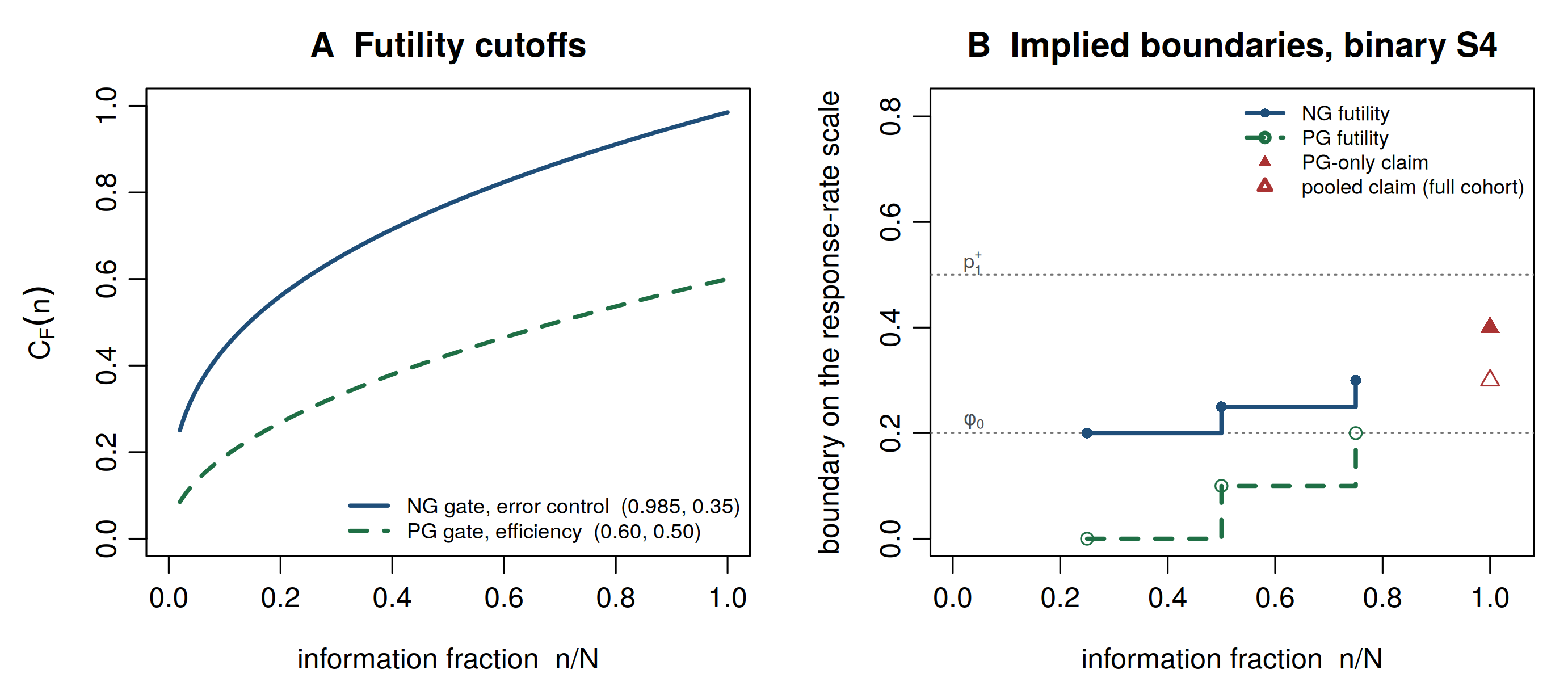}
\caption{Futility cutoffs and decision boundaries for binary scenario S4. \textbf{A} NG and PG futility cutoffs, $C_F(n) = \lambda(n/N)^{\gamma}$. \textbf{B} Corresponding count boundaries expressed as response proportions. The pooled boundary applies only at the final analysis after both cohorts reach their planned sizes.}\label{fig1}
\end{figure}

As an illustration, Figure~\ref{fig1} shows the calibrated cutoff functions and the integer boundaries they imply for a single binary endpoint. The tabulated boundaries reproduce the posterior decisions exactly in every scenario and endpoint (Appendix S3).

\section{Simulation Study}\label{sec3}

\subsection{Scenarios, look schedule, and calibration settings}\label{sec3.1}

As a concrete setting, consider a single-arm phase II trial of a targeted agent in a previously treated solid tumor, planned as scenario S4 below. A molecular alteration present in roughly one third of patients is thought to predict benefit, but the mechanism does not exclude activity in unselected patients, and the sponsor wishes to retain the option of an all-comer claim. Historical objective response rates under standard therapy are about 20\% irrespective of biomarker status; a response rate of 40\% in biomarker-negative and 50\% in biomarker-positive patients would justify further development, so the trial enrolls 40 biomarker-negative and 20 biomarker-positive patients with three interim looks at equal information fractions (after 10, 20 and 30 biomarker-negative and 5, 10 and 15 biomarker-positive patients). Because response is assessed at the first post-baseline scan, roughly eight weeks after enrollment, the two cohorts can be reviewed at synchronized looks with a short accrual pause. The protocol states the integer rules of Table~\ref{tab2}: at the first look the biomarker-negative cohort is frozen if at most 2 of its first 10 patients have responded, provided that the biomarker-positive cohort has not met its own futility criterion (in which case the trial stops); enrollment then continues in the biomarker-positive subgroup alone, and a positive result there requires at least 8 responses among its 20 patients. If both cohorts pass all three looks, a claim for the whole population requires at least 18 responses among the 60 patients. Two consequences of the design are visible at the planning stage: the exact upper bound on a false joint claim is 0.045 (Table~\ref{tab3}), and, because the same gate caps the joint-claim power at $P(S^{\ast}) = 0.72$ for this biomarker-negative cohort (Table~\ref{tabS14}), a sponsor for whom the all-comer claim is the primary objective would need a larger biomarker-negative cohort (Figure~\ref{fig3}), whereas a sponsor targeting the biomarker-positive claim can accept the trade-off. The example is illustrative and does not describe a specific product; the same construction applies to the other scenarios.

BOP2-ENR was evaluated across eight response-rate scenarios used to assess the BOOST design, four symmetric ($\phi_0^- = \phi_0^+$; S1--S4) and four asymmetric ($\phi_0^- \ne \phi_0^+$; S5--S8), written here in the notation of Section~\ref{sec2}, with maximum sample sizes rounded to convenient values close to the published optimal weak-control designs; the maxima are 120, 80, 60 and 40 patients. The rounded totals differed from the published BOOST designs by at most nine patients, and both were reported in the comparison tables. In the asymmetric scenarios the pooled null threshold $\phi_0^{\mathrm{pool}}$ lies strictly between the two subgroup nulls, so that the pooled claim tests neither subgroup's hypothesis but a weighted compromise between them. The same eight settings, sample sizes and look schedule were used for all three endpoint structures, so that differences between the endpoint sections that follow are attributable to endpoint structure alone. The scenario parameters specific to each endpoint are tabulated at the head of its section.

All operating characteristics were evaluated with $10^5$ replicates; BOP2-ENR calibration scans used $3 \times 10^4$ replicates, whereas the BOP2-P cutoff was calibrated at $10^5$ (its reported $\mathrm{FWER}_{00}$ being the value at selection); the look-schedule sensitivity analysis of Appendix S2 used $2.5 \times 10^4$ for calibration and $6 \times 10^4$ for evaluation. The nuisance parameter $\psi$ was profiled over a three-point grid on the path interpolating from the PG null vector to the extreme alternative, at interpolation fractions $t = 0.3$, $0.65$ and $1$, with $t = 1$ the extreme alternative in which the favorable categories carry nearly all mass; a three-point grid suffices because $\mathrm{FWER}_{01}(\psi)$ is non-decreasing along this path and reaches its plateau early (Section~\ref{sec4.0}, Appendix S1). For the multi-criterion endpoints the two component outcomes were taken independent given their marginals. This is a design assumption: the marginal null thresholds do not determine the four-category null vector, so the calibration was carried out under a prespecified joint null distribution, and the strong-control guarantee is stated for that distribution. Its sensitivity to association was examined in Table~\ref{tabS15}. Under independence the four categories are ordered by the presence of the first and second outcome, (1,1), (1,0), (0,1), (0,0), so the first criterion aggregates categories 1 and 2 ($\mathbf{b} = (1,1,0,0)$) and the second categories 1 and 3 ($\mathbf{b} = (1,0,1,0)$); for efficacy-toxicity the second outcome is toxicity, whose presence is unfavorable, and the ordering is $(E,T) = (1,1), (1,0), (0,1), (0,0)$. The nominal one-sided level was $\alpha = 0.05$; at $10^5$ replicates the Monte Carlo standard error of an error-rate estimate near $\alpha$ was about 0.0007. Four looks ($R = 4$) were used throughout at equally spaced information fractions: three interim futility analyses and a final analysis. Relative to $R = 4$, a single interim analysis ($R = 2$) increased the mean expected sample size under the global null by 6.4 to 7.0 patients for the binary and efficacy-toxicity endpoints and was feasible in only five of eight co-primary scenarios, whereas $R = 6$ reduced it by a further 2.1 to 2.9 patients; the sensitivity analysis over $R = 2, \dots, 6$ is reported in Appendix S2 and Tables~\ref{tabS1}, \ref{tabS4} and \ref{tabS7}.

\subsection{Comparators}\label{sec3.2}

Three reference designs were evaluated on the same settings. For each, the certifying bound $P(S)$ of Proposition (a) was evaluated on its own NG futility rule: for BOOST it is the exact probability of passing the Stage 1 NG boundary, whereas BOP2-P and BOP2-ENR-nE never freeze the NG cohort, so their bound is one.

\textbf{BOOST.} The published two-stage designs \citep{Tong2026-es}, evaluated exactly by enumeration rather than by simulation. In the Simon tradition each design is published in an \textit{optimal} form, which minimizes the expected sample size under the null, and a \textit{minimax} form, which minimizes the maximum sample size; the optimal designs are used here, as the efficiency criterion relevant to the expected-sample-size comparison. BOOST is a biomarker-guided Simon two-stage design: a Stage 1 futility look routes the trial between enrichment to the biomarker-positive subgroup and unselected continuation, and a Stage 2 decision covers pooled and subgroup-specific claims. It is published in two variants, distinguished by the multiplicity criterion its optimization enforces. The \textit{weak-control} variant bounds the family-wise error rate only under the global null, where every subgroup is inactive; it is the more efficient of the two but does not constrain the error rate when one subgroup is active and the other is not. The \textit{strong-control} variant additionally bounds the error rate across the mixed configurations, at the cost of a larger sample size. Both variants were included because the contrast between them is exactly the weak-versus-strong distinction that BOP2-ENR is designed to address, and comparing against each separates the efficiency cost of the stricter criterion from the cost of the endpoint generalization. BOOST is defined for a binary endpoint only, so it appears in Table~\ref{tab3} alone.

\textbf{BOP2-P.} The original BOP2 design \citep{Zhou2017-tv} applied to the unselected population: a single cohort of size $N^- + N^+$ drawn from the biomarker mixture, futility monitoring only at the conventional tuning ($\lambda = 0.60$, $\gamma = 0.50$), and a single claim about the overall population. This isolated the cost of ignoring the biomarker entirely. BOP2-P used the same maximum total sample size and the planned mixture weight $w = N^-/(N^- + N^+)$ as BOP2-ENR, but drew every patient from the mixture $w\theta^- + (1-w)\theta^+$ rather than fixing the subgroup counts, so its sampling distribution differed from that of the fixed-quota designs. Its efficacy cutoff was calibrated to the same nominal level against its own null: the pooled null point for the single binary and co-primary endpoints, and for joint efficacy-toxicity monitoring both faces of the pooled union null (the efficacy-null face at its least-favorable toxicity rate, and the toxicity-null face at pooled efficacy one), the reported $\mathrm{FWER}_{00}$ being the larger of the two. BOP2-P targets a mixture-level claim. For the present comparison we additionally evaluated the probability that this claim would be interpreted as evidence of benefit in both subgroups when NG is inactive: $\max_{\mathrm{grid}}\mathrm{FWER}_{01}$ for BOP2-P is the profiled probability of any claim under the mixed configuration. This is an error measure under the joint-claim interpretation used here, not the type I error rate of BOP2-P for its own mixture estimand; BOP2-ENR-nE, which keeps the claim architecture of BOP2-ENR, is the more direct comparator.

\textbf{BOP2-ENR-nE.} An ablation of the proposed design in which the enrichment route is disabled: an NG futility signal no longer freezes NG accrual, so both cohorts accrue to their maxima unless PG signals futility. All tuning parameters were held at their calibrated BOP2-ENR values, so the comparison isolated the contribution of enrichment.

\subsection{Strong control and the boundary of the nuisance path}\label{sec4.0}

\begin{figure}[htbp]
\centering
\includegraphics[width=\textwidth]{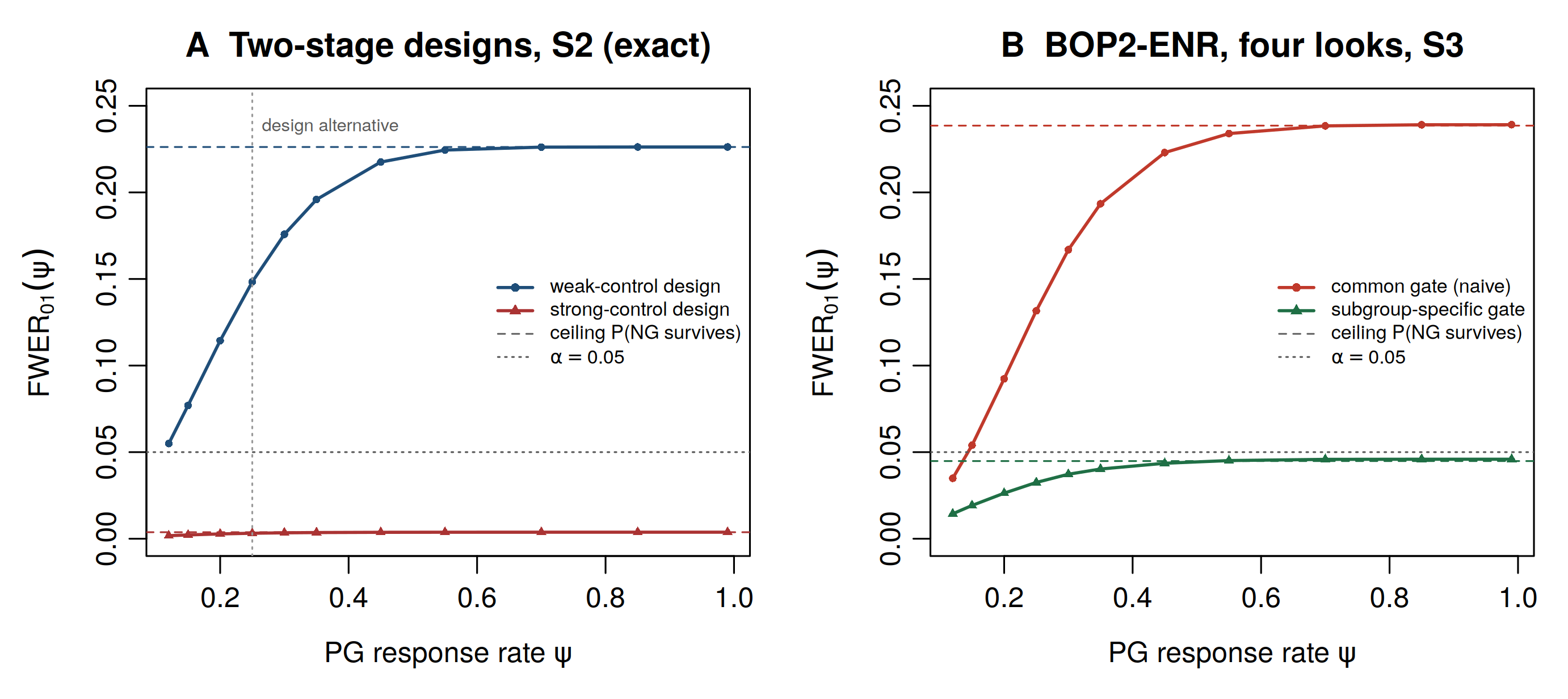}
\caption{Mixed-configuration error profiles and their upper bounds. Profiles of $\mathrm{FWER}_{01}(\psi)$ for \textbf{A} the published BOOST weak- and strong-control designs in S2, evaluated exactly, and \textbf{B} four-look BOP2-ENR with common or subgroup-specific futility tuning in S3. Dashed lines indicate the corresponding NG-survival bounds, dotted lines the nominal level, and vertical lines the design alternatives.}\label{fig2}
\end{figure}

Profiling $\mathrm{FWER}_{01}$ over the nuisance parameter for the published BOOST weak-control design \citep{Tong2026-es} (scenario S2) showed a monotone rise to a plateau: 0.0413, 0.1144, 0.1483, 0.1959, 0.2221, 0.2262, 0.2262, 0.2262 at $\psi = 0.10, 0.20, 0.25, 0.35, 0.50, 0.70, 0.90, 0.99$ (Figure~\ref{fig2}A). The value at the design alternative is 0.1483, but the supremum is 0.2262; evaluating at the design alternative alone would understate the error rate by a third. The supremum coincides exactly with the ceiling of Proposition (a): the published S2 weak-control design enrolls $n_1^- = 5$ NG patients in Stage 1 and continues NG if at least one responds, so under $\phi_0^- = 0.05$ the NG cohort survives with probability $P(\mathrm{Bin}(5, 0.05) \ge 1) = 0.2262$.

The same mechanism separates two calibrations of BOP2-ENR. With a common futility tuning ($\lambda = 0.60$, $\gamma = 0.50$) in both subgroups, weak control holds but $\mathrm{FWER}_S = 0.239$ in binary scenario S3, equal to the computed ceiling to three decimal places (Figure~\ref{fig2}B); with subgroup-specific tuning $\mathrm{FWER}_S = 0.046$. The difference lies in the NG gate alone: posterior gates at conventional values are more permissive than the count gates of the two-stage designs, and because the bound does not involve the efficacy boundary, the mixed error is not corrected by the efficacy cutoff. Weak-control evaluation alone would not detect this.

The ceiling responds monotonically to the NG gate: in binary scenario S3 with the PG gate fixed and $\gamma_- = 0.35$, NG cutoffs $\lambda_- = 0.60, 0.80, 0.90, 0.95$ gave ceilings 0.239, 0.104, 0.079 and 0.045.

For co-primary endpoints in S3, varying $\lambda_E$ over 0.96, 0.97 and 0.98 left $\max_{\mathrm{grid}}\mathrm{FWER}_{01}$ at 0.041 while $\mathrm{FWER}_{00}$ fell from 0.086 to 0.030 (Table~\ref{tabS13}), consistent with the Remark of Appendix S1: within the evaluated range the profiled mixed-configuration error was nearly invariant to the efficacy cutoff, whereas the global-null error changed substantially. The bound was near-attained beyond the binary case as well: in every scenario-by-endpoint combination the reported profiled maximum (seed 77) and the corresponding NG-survival probability $P(S)$ agreed to within 0.0015 (within 0.0008 on independent re-evaluation), the order of the Monte Carlo error of the profiled estimate, the bound itself being exact (Table~\ref{tabS12}).

In every scenario-by-endpoint combination, $\mathrm{FWER}_{01}(\psi)$ was flat or nearly flat over the upper part of the profiling path, consistent with the ceiling being attained well before the extreme alternative. For the binary endpoint the profile was already at its plateau at the first grid point in most scenarios, which is what makes the three-point grid of Section~\ref{sec3.1} sufficient.

\subsection{Single binary endpoint}\label{sec4.1}

Table~\ref{tab1} gives the scenario parameters of Section~\ref{sec3.1} and Table~\ref{tab2} the rules a protocol would state, obtained from the calibrated tuning parameters (Table~\ref{tabS2}) by the inversion of Section~\ref{sec2.6} and verified against the decisions taken by the simulator (Appendix S3). The NG futility cutoff was uniformly aggressive because it alone bounds the mixed-configuration component of the restricted strong FWER, the final efficacy cutoff being calibrated to control the global-null component, while the PG cutoff was permissive because it served efficiency.

Table~\ref{tab3} compares BOP2-ENR with the reference designs of Section~\ref{sec3.2} for the single binary endpoint. BOP2-ENR controlled the strong FWER in all eight scenarios, with the exact bound $P(S)$ between 0.024 and 0.046 and the profiled diagnostic $\mathrm{FWER}_S$ between 0.024 and 0.047. $\mathrm{POWER}_{01}$ reached 0.80 or above in four of eight (0.75 or above in seven), and disjunctive power exceeded the BOOST weak-control design in seven of eight scenarios. The second configuration-specific criterion, $\mathrm{POWER}_{11}$ (the probability of claiming for both subgroups when both are active) ranged from 0.43 to 0.69, well below $\mathrm{POWER}_{01}$ in all scenarios. The Corollary of Section~\ref{sec2.5} quantifies why: the NG gate that bounds the error rate also freezes the NG cohort early.

\textbf{The profiled error rate separated the designs sharply.} For BOP2-ENR, $\max_{\mathrm{grid}}\mathrm{FWER}_{01}$ never exceeded 0.047. For the BOOST weak-control designs it ranged from 0.024 to 0.226 and in every scenario equalled, to four decimal places, the probability that the NG cohort passes its Stage 1 futility boundary; the BOOST strong variants brought it to 0.004--0.050 through their NG Stage 1 gate.

\textbf{Under the joint-subgroup interpretation pooled BOP2 had a high mixed false-claim probability, and the ablation lost restricted strong control.} BOP2-P controlled the error rate of the single claim it makes (0.025--0.039) but, having no subgroup-restricted alternative, made a claim in 49--71\% of replicates when only PG was active ($\max_{\mathrm{grid}}\mathrm{FWER}_{01}$ of 0.988--1.000); under the joint-claim interpretation of Section~\ref{sec3.2} this is not addressed by calibrating the final efficacy cutoff against the global null, because it concerns a different configuration. Disabling only the enrichment route, with every tuning parameter at its calibrated value, raised $\max_{\mathrm{grid}}\mathrm{FWER}_{01}$ to 1.000 in all eight scenarios and $\mathrm{ESS}_0$ by 6--35 patients, consistent with Proposition (a): without enrichment the NG cohort always survives, the bound is one, and the simulated mixed-configuration error approached one in these settings. Removing the gate also raised the global-null error rate, to 0.046--0.069 for the binary endpoint, because the pooled claim, which is available only when both cohorts complete, becomes available whenever the trial reaches the final analysis without stopping for PG futility: with the tuning parameters calibrated for the enriched design, the ablation's Monte Carlo estimate of the global-null error rate exceeded 0.05 in seven of eight binary scenarios and in 17 of the 24 scenario-by-endpoint combinations (Tables~\ref{tab3}, \ref{tab6} and \ref{tab9}); estimates close to 0.05, such as 0.0503 in binary S2 and S4, lie within Monte Carlo uncertainty of the nominal level. The NG gate therefore contributes to both error components and directly calibrates the uniform upper bound on the mixed-configuration error, which does not depend on the efficacy cutoff or on the PG gate.

\textbf{Expected sample size.} Three comparisons are reported as found. Against the published BOOST weak-control rows, BOP2-ENR required a larger $\mathrm{ESS}_0$ in six of eight scenarios (approximately equal in S3, smaller in S7; mean difference $+7.9$ patients), the price of meeting strong rather than weak control and of graduated posterior boundaries in place of a small hard-gated first stage. Against the published BOOST strong variants it was more efficient in one of seven, but five of those seven exceed $\alpha$ in $\mathrm{FWER}_{00}$ under exact evaluation (flagged in Table~\ref{tab3}) and are not a like-for-like benchmark. Against the BOOST designs locally re-optimized under the stated exact pointwise constraints (Table~\ref{tabS11}; a $\pm 3$ re-search, so indicative rather than optimal), BOP2-ENR had the smaller $\mathrm{ESS}_0$ in five of seven (S1--S5), which suggests that much of the gap against the published rows reflects the error-control criterion rather than the endpoint generalization, although sample sizes, look schedules, boundary families and the local search range also differ between the designs.

\subsection{Co-primary endpoints}\label{sec4.2}

The scenarios were those of Table~\ref{tab1} with a second criterion, the PFS6 rate, added (Table~\ref{tabScp}); response and PFS6 were taken independent given their marginals. For this endpoint the choice of look schedule was more constrained than for the others: with the prespecified sample sizes, tuning grid and bound-based calibration criterion, no feasible candidate was found for three of the eight scenarios with a single interim analysis ($R = 2$; Section~\ref{sec3.1} and Appendix S2), whereas the four-look schedule adopted here was feasible in all eight.

In the boundaries of Table~\ref{tab5}, futility required both criteria to be crossed at the same look, so that a trial continued while either criterion remained promising; the tuning parameters generating these rules are given in Table~\ref{tabS5}.

BOP2-ENR performed strongly for co-primary endpoints (Table~\ref{tab6}), with $\mathrm{POWER}_{01} \ge 0.75$ in all scenarios (0.80 or above in seven of eight) and disjunctive power between 0.89 and 0.98. $\mathrm{POWER}_{11}$ ranged from 0.39 to 0.74, as for the binary endpoint and for the same structural reason (the Corollary of Section~\ref{sec2.5}). The disjunctive efficacy rule gave two routes to a claim, and the additional PFS6 information was substantial when the response rate was low.

The price appeared in the calibrated cutoffs of Table~\ref{tabS5}. Relative to the binary endpoint the efficacy cutoff was displaced upward, to 0.970--0.985 against 0.88--0.95, because the disjunctive rule across two criteria inflates the type I error rate, the phenomenon BOP2 absorbs through cutoff calibration \citep{Zhou2017-tv}, here compounded by the pooled-versus-subgroup multiplicity. The NG gate was displaced in shape rather than level: $\gamma_-$ fell to 0.10 in six scenarios and 0.20 in S1 and S8, against 0.35 for the other two endpoints, while $\lambda_-$ remained in the same range (0.92--0.995). A smaller $\gamma_-$ raises the cutoff at the early looks, as the conjunctive futility rule requires, since demanding that both criteria fail makes the NG cohort harder to stop.

The comparators behaved as in the binary case. BOP2-P again controlled its own type I error rate and attained very high disjunctive power (0.95--0.99 in most scenarios, since two criteria gave it two routes to a claim as well) while overall efficacy was asserted in 55--82\% of replicates when only PG was active, with $\max_{\mathrm{grid}}\mathrm{FWER}_{01}$ above 0.99 throughout. The no-enrichment ablation again lost strong control completely, and its Monte Carlo estimate of the global-null error rate exceeded 0.05 in five of eight scenarios (0.033--0.065). Notably, the efficiency gap between BOP2-ENR and the ablation was larger here than for the binary endpoint, since the conjunctive futility rule made early stopping harder and the frozen NG cohort saved correspondingly more patients.

\subsection{Joint efficacy-toxicity monitoring}\label{sec4.3}

{The scenarios were those of Table~\ref{tab1} with a toxicity criterion added (Table~\ref{tabSet}). The null toxicity threshold was fixed at $\phi_{02} = 0.30$ and the tolerable (true) toxicity rate at 0.10 in every scenario, in both subgroups. Because the subgroup null for this endpoint is the union of the efficacy-null and toxicity-null faces (Section~\ref{sec2.2}), strong control was evaluated on both: the NG gate was calibrated against the exact bound at $(\phi_{01}, 0)$, the safety cutoffs $\lambda_T^{\pm}$ against the exact size at $\theta_T = 0.30$ for $N^-$ and $N^+$, and $\mathrm{FWER}_{00}$ was maximized over the four admissible null corners.

In the boundaries of Table~\ref{tab8}, futility required either criterion to be crossed while a claim required both, the toxicity rows carrying the opposite sense and safety being established separately in each cohort (tuning parameters in Table~\ref{tabS8}). The NG gate sat exactly where the binary calibration placed it ($\gamma_- = 0.35$ throughout, $\lambda_-$ from 0.92 to 0.985), because on the efficacy-null face the toxicity criterion never trips and the gate is the binary gate; the safety cutoffs were $\lambda_T = 0.95$ at $N = 20$ and $0.97$ at $N = 40$ to $80$, the smallest grid values with exact size at most 0.05 at the toxicity threshold.

Strong control held for BOP2-ENR on both faces in all scenarios (Table~\ref{tab9}): the certifying bound was between 0.034 and 0.046, the toxicity face contributing at most 0.034, and the profiled diagnostic $\mathrm{FWER}_S$ was between 0.036 and 0.048, with $\mathrm{FWER}_{00}$ maximized over the four admissible null corners. Power was materially lower than for the other endpoints, with $\mathrm{POWER}_{01}$ between 0.51 and 0.81 (0.80 or above only in S6, whose PG cohort is the largest at 60) and $\mathrm{POWER}_{11}$ between 0.20 and 0.48. The cause was the safety criterion itself rather than the enrichment machinery. A claim must establish, within the claiming cohort, that the toxicity rate is below 0.30 with posterior probability $\lambda_T$ calibrated so that a cohort at the threshold passes with probability at most 0.05; at a true rate of 0.10 the probability of passing is 0.68 at $N = 20$, 0.90 at $N = 40$ and 0.99 at $N = 60$, and the PG-claim power is essentially the binary power multiplied by this factor. The joint claim requires safety in both cohorts and pays the factor twice, which is why its refined ceiling in Table~\ref{tabS14} lies well below $P(S^{\ast})$.

The same limitation affected BOP2-P, whose disjunctive power was likewise depressed relative to the other endpoints (0.76--0.95, with $\mathrm{POWER}_{01}$ of 0.38--0.70) once its single cutoff was calibrated to protect the toxicity face of its own union null as well as the efficacy face, which required $\lambda_E = 0.96$--$0.97$ against 0.88--0.96 for the binary endpoint, confirming that the constraint was the endpoint structure and the sample size rather than the enrichment machinery. The Monte Carlo estimate of the no-enrichment ablation's global-null error rate exceeded 0.05 in five of eight scenarios (0.047--0.076). Trials with a conjunctive efficacy-toxicity claim require cohorts large enough to establish safety at the nominal level, of the order of 40 patients per claiming cohort at the rates used here, and a design intended for this endpoint should be sized for it rather than inherit the sample sizes of a single binary endpoint.}

\subsection{Verification and reproducibility}\label{sec3.3}

Verification of the computational engines against exact enumeration, of the tabulated boundaries against the decisions taken by the simulator, and the exact recomputation of the published BOOST operating characteristics \citep{Tong2026-es} are reported in Appendix S3; the BOOST rows of the comparison tables are our exact recomputation at the published design parameters rather than the published rates. The calibrated tuning parameters that generated the boundaries, and the distribution of the look at which enrichment occurred, are given in the Supporting Information.

{Because $\mathrm{FWER}_{00}$ is estimated by Monte Carlo whereas $P(S)$ is exact, $U_S$ has two versions. The point-estimate version $\widehat{U}_S = \max\{\widehat{\mathrm{FWER}}_{00}, P(S)\}$ is reported as a column of the comparison tables (Tables~\ref{tab3}, \ref{tab6} and \ref{tab9}) and summarized in Table~\ref{tabsum}. The verification version $U_S^{\mathrm{UCL}} = \max\{\mathrm{UCL}_{00}, P(S)\}$ replaces the estimate by a one-sided 95\% Clopper-Pearson upper confidence limit, Bonferroni-adjusted for the number of Monte Carlo maximands behind the reported maximum (the four null corners for efficacy-toxicity; one otherwise); it is reported as the last column of Table~\ref{tabS12}, which also lists $P(S)$, the adjusted limit of $\mathrm{FWER}_{00}$ and, for the profiled diagnostic, the corresponding limit adjusted over its faces and grid points. Every selected design satisfied $U_S^{\mathrm{UCL}} \le \alpha$. The adjustment is within a design; across designs, the single Bonferroni limit over all 48 global-null corner estimates is 0.0482, and no further simultaneous guarantee is claimed. The mixed-configuration component is thus controlled by an exact upper bound, whereas the global-null component was calibrated by Monte Carlo with a prespecified margin and independently re-evaluated with multiplicity-adjusted upper confidence limits.}

\section{Discussion}\label{sec5}

{The aim was to add adaptive enrichment and pooled/subgroup multiplicity control to BOP2 while keeping its endpoint generality, flexible interim schedule and posterior-scale decision rules. The construction met the prespecified exact-bound and Monte Carlo verification criteria for strong FWER control in all scenario-by-endpoint combinations: the mixed-configuration component is controlled by an exact upper bound, the probability that the biomarker-negative cohort survives its own futility monitoring, and the global-null component was calibrated by Monte Carlo with a prespecified margin and independently re-evaluated with multiplicity-adjusted one-sided upper confidence limits, the largest of which, simultaneous over all 48 global-null corner estimates, is 0.048 (Table~\ref{tabS12}). Strong control is claimed on a restricted parameter space: the monotone-activity assumption, which excludes the configuration in which only the biomarker-negative subgroup is active, and the component-level efficacy ordering $\theta_E^- \le \theta_E^+$. Both are substantive. In the asymmetric scenarios the excluded direction gives false PG claim rates of 0.070 to 0.161 (Table~\ref{tabS10}), because a raw-rate ordering alone cannot exclude it there, and without the efficacy ordering the global null of the efficacy-toxicity endpoint admits a corner at which the design's any-claim probability was 0.06 to 0.07; checking both premises against external evidence is a precondition for adopting the design.

The central structural result is that, in any design where a subgroup claim is available only jointly with another's through a pooled analysis, the NG futility gate governs both faces of the design. It bounds the mixed-configuration error rate by $P(S)$ under the NG null and, by the same event inclusion, bounds the joint power by $P(S^{\ast})$ under the NG alternative; calibrating the gate to $P(S) \le \alpha$ at NG sizes of 20 to 80 patients leaves $P(S^{\ast})$ at 0.36 to 0.74, which $\mathrm{POWER}_{11}$ attained to within 0.023 for co-primary endpoints and to within 0.104 for the binary endpoint (Table~\ref{tabS14}). The gate is not the only determinant of the global-null component, but it enters there too: with the calibrated cutoffs left in place, removing it produced Monte Carlo estimates of the global-null error rate above the nominal level in 17 of 24 settings, because the pooled claim then becomes available whenever the trial reaches the final analysis without stopping for PG futility. The bound does not involve the efficacy cutoff, so tightening the efficacy cutoff cannot be relied on to restore strong control, and a design calibrated at conventional alternative points can satisfy weak control while violating strong control by nearly fivefold; nuisance-path profiling is a tightness diagnostic and should accompany, not replace, the bound. Joint power, correspondingly, is not a calibration failure to be repaired by the efficacy cutoff or the PG gate: the NG gate places an upper bound on joint-claim power at the planned NG sample size, and tightening the final efficacy cutoff or modifying the PG gate cannot raise this ceiling, although these choices affect the power attained below it. Relaxing the NG gate trades a higher power ceiling against a larger mixed-error bound, and this trade-off need not be one-for-one; a design whose primary objective is the joint claim should therefore be sized for the NG cohort (Figure~\ref{fig3}). Decoupling the two would require subgroup-specific NG evidence in addition to pooled significance, since pooled significance is a decision-level joint claim and not a posterior demonstration of efficacy in NG, at the cost of the pooled rule that gives the design its simplicity. Three practical recommendations follow: treat the NG futility boundary as an error-control parameter tuned separately from its PG counterpart, which serves efficiency; use the single-cohort bound as the calibration target, which decouples an otherwise five-dimensional search; and report the bound and the profiled diagnostic side by side.

The endpoint structures differed in instructive ways. Co-primary endpoints, whose disjunctive claim rule gives two routes to a claim, required a higher efficacy cutoff and a reshaped NG gate but delivered the highest power. Joint efficacy-toxicity monitoring taught a lesson the other endpoints could not: when a subgroup can be inactive for either of two reasons the null is a union, all four combinations of the two faces belong to the global null, and a pooled statistic for the second criterion is unsafe in both directions, since a safe cohort masks a toxic one exactly as an effective cohort masks an ineffective one. Establishing safety within each subgroup with an exactly calibrated cutoff controls both faces at a cost in power that is the power of the safety criterion itself at the cohort size, about 0.68 at 20 patients and 0.90 at 40 for a tolerable rate of 0.10 against a threshold of 0.30. We regard this as the correct price: a joint claim of efficacy and safety for a 20-patient subgroup should not be easy to make, and a trial with this endpoint should be sized for its safety criterion rather than inherit the sample sizes of a single binary endpoint.

The comparison with two-stage designs does not support a claim of uniform superiority, and the expected sample size comparison is reported as found. Against the published weak-control BOOST designs BOP2-ENR was less efficient in six of eight binary scenarios while delivering higher disjunctive power in seven of eight and meeting a stricter error-control criterion; the published strong variants are not a like-for-like benchmark, since five of seven exceed $\alpha$ under exact evaluation, and against the designs locally re-optimized under the stated exact pointwise constraints BOP2-ENR was the more efficient in five of seven (Table~\ref{tabS11}). BOP2-ENR is applicable in settings not covered by the evaluated comparator, namely co-primary and efficacy-toxicity endpoints and schedules with more than one interim look, where no relative efficiency claim is made.

Several limitations bear on interpretation. Pooling is fixed-effect: the pooled claim is evaluated with a single working Beta model for the allocation-weighted mixture, so that the two subgroups are combined with fixed weights and no strength is borrowed adaptively between them. A hierarchical or exchangeability-based pooling would change the pooled statistic and its calibration, but not the event-inclusion argument, since the NG futility gate is unaffected. The co-primary and efficacy-toxicity designs are calibrated under independence between the component outcomes, a design assumption rather than a nuisance parameter: over odds ratios of 0.25 to 4 the certified bounds and the corner-maximized global-null error rates stayed at or below 0.05 in point estimate, but power moved by up to 0.10 and one adjusted upper confidence limit slightly exceeded 0.05 (Tables~\ref{tabS15} and \ref{tabS16}). Sample sizes were rounded values close to published two-stage designs rather than optimized per endpoint, which particularly disadvantages the efficacy-toxicity endpoint. Calibration grids were coarse and the profiling path one-dimensional; attainment of the bound at the boundary of the path was established analytically for the binary and co-primary endpoints (Appendix S1) but only observed numerically for efficacy-toxicity. Extensions that leave the futility gate intact, such as time-to-event endpoints \citep{Zhou2020-gh,Lin2020-wz} or dual-criterion decision making \citep{Zhao2023-hu}, inherit the bound unchanged and the sequential calibration of Section~\ref{sec2.5b} should carry over with the Beta tail replaced by the corresponding posterior quantity; extensions that add a route to a claim, such as efficacy stopping \citep{Xu2025-xq}, require the two stopping rules to be calibrated jointly, and a randomized version \citep{Zhao2022-xr} would redefine the mixed configuration in terms of treatment effects without changing the structural argument.}

\section*{Conflict Of Interest Statement}

The authors have declared no conflict of interest.

\section*{Declaration of Generative AI Use}

Claude Fable 5.1 (Anthropic) was used to assist with language editing of the manuscript and with code review. The authors verified all generated suggestions and take full responsibility for the manuscript and the reported results.

\section*{Code and Data Availability Statement}

This is a simulation study and involved no patient data. The analysis code (base R), seeds, cached outputs and scripts for regenerating the figures and the numerical bodies of the 24 code-generated tables are supplied as a Data File with this submission; Tables 1, S4 and S10 are manually prepared configuration and scenario summaries. The design code will be released as an R package on CRAN (bop2enr) and in a public GitHub repository (\url{https://github.com/belayb/BOP2-ENR}) upon acceptance.

\bibliographystyle{wileyNJD-AMA}
\bibliography{paperpile}

\newpage
\newpage

\newpage
\begin{table}[htbp]
\centering
\caption{Subgroup activity configurations and claim interpretation.}\label{tab0}
{\fontsize{7.5pt}{9pt}\selectfont\setlength{\tabcolsep}{2.5pt}
\begin{tabular}{llll}
\toprule
\textbf{Configuration} & \textbf{True state} & \textbf{Erroneous decision(s)} & \textbf{Correct decision(s)} \\
\midrule
$\Theta_{00}$: $\theta^- = \theta_0^-,\ \theta^+ = \theta_0^+$ & both subgroups null & PG-restricted claim; pooled claim & no claim \\
$\Theta_{01}$: $\theta^- = \theta_0^-,\ \theta^+ = \psi \in \Theta_1^+$ & NG null, PG active & pooled claim (false NG claim) & PG-restricted claim \\
$\Theta_{11}$: $\theta^- = \theta_1^-,\ \theta^+ = \theta_1^+$ & both subgroups active & --- & pooled claim; PG-restricted claim \\
$\Theta_{10}$: $\theta^- \in \Theta_1^-,\ \theta^+ = \theta_0^+$ & NG active, PG null & (excluded by assumption) & --- \\
\bottomrule
\end{tabular}}
\tabnote{A pooled rejection is interpreted as a joint NG--PG claim and is available only on the continue-both path; a PG-restricted claim is available after enrichment or as a fallback (Section~\ref{sec2.3}). $\Theta_{10}$ is excluded by the monotone-activity assumption (Table~\ref{tabS10}). Efficacy-toxicity nulls comprise efficacy and toxicity faces (Section~\ref{sec2.4}). $\theta_0^{\pm}$ denotes the prespecified null point of a subgroup; $\Theta_{00}$, $\Theta_{01}$ and $\Theta_{11}$ are configurations of the truth, displayed at representative points, with null configurations evaluated at their least-favorable points.}
\end{table}
\newpage
\begin{table}[htbp]
\centering
\caption{Scenario parameters and subgroup sample sizes, single binary endpoint.}\label{tab1}
{\fontsize{9pt}{10.8pt}\selectfont
\begin{tabular}{lccccc}
\toprule
\textbf{Sc.} & \textbf{Null $(p_0^-,p_0^+) = (\phi_0^-,\phi_0^+)$} & \textbf{Alt. $(p_1^-,p_1^+)$} & \textbf{$N^-$} & \textbf{$N^+$} & \textbf{Pooled null $\phi_0^{\mathrm{pool}}$} \\
\midrule
S1 & (0.03, 0.03) & (0.10, 0.15) & 80 & 40 & 0.030 \\
S2 & (0.05, 0.05) & (0.20, 0.25) & 20 & 20 & 0.050 \\
S3 & (0.10, 0.10) & (0.25, 0.35) & 40 & 20 & 0.100 \\
S4 & (0.20, 0.20) & (0.40, 0.50) & 40 & 20 & 0.200 \\
S5 & (0.05, 0.10) & (0.20, 0.25) & 20 & 40 & 0.083 \\
S6 & (0.05, 0.20) & (0.20, 0.35) & 20 & 60 & 0.163 \\
S7 & (0.05, 0.10) & (0.25, 0.30) & 20 & 20 & 0.075 \\
S8 & (0.10, 0.20) & (0.30, 0.45) & 20 & 20 & 0.150 \\
\bottomrule
\end{tabular}}
\tabnote{$p_0^{\pm}$ and $p_1^{\pm}$ are subgroup null and alternative response rates; $\phi_0^{\mathrm{pool}} = (N^- p_0^- + N^+ p_0^+)/(N^- + N^+)$.}
\end{table}
\newpage
\begin{table}[htbp]
\centering
\caption{Decision boundaries for the binary endpoint.}\label{tab2}
{\fontsize{9pt}{10.8pt}\selectfont
\begin{tabular}{lccccc}
\toprule
\textbf{Sc.} & \textbf{Criterion} & \textbf{Futility, NG} & \textbf{Futility, PG} & \textbf{PG-only claim} & \textbf{Pooled claim} \\
\midrule
S1 & ORR & $\le$ 1, 2, 3 & $\le$ ---, 0, 0 & $\ge$ 4 & $\ge$ 7 \\
S2 & ORR & $\le$ 0, 1, 2 & $\le$ ---, ---, 0 & $\ge$ 4 & $\ge$ 5 \\
S3 & ORR & $\le$ 1, 3, 5 & $\le$ 0, 0, 1 & $\ge$ 5 & $\ge$ 10 \\
S4 & ORR & $\le$ 2, 5, 9 & $\le$ 0, 1, 3 & $\ge$ 8 & $\ge$ 18 \\
S5 & ORR & $\le$ 0, 1, 2 & $\le$ 0, 0, 1 & $\ge$ 8 & $\ge$ 9 \\
S6 & ORR & $\le$ 0, 1, 2 & $\le$ 1, 4, 7 & $\ge$ 18 & $\ge$ 19 \\
S7 & ORR & $\le$ 0, 1, 2 & $\le$ 0, 0, 0 & $\ge$ 5 & $\ge$ 6 \\
S8 & ORR & $\le$ 0, 1, 3 & $\le$ ---, 0, 0 & $\ge$ 8 & $\ge$ 11 \\
\bottomrule
\end{tabular}}
\tabnote{Futility counts correspond to information fractions 0.25, 0.50 and 0.75; the displayed inequalities define stopping and final claims. A dash denotes an unattainable futility boundary. PG futility stops the trial; NG futility triggers enrichment if PG continues. The pooled claim requires both cohorts to reach their planned sizes.}
\end{table}
\newpage
\begin{center}\setlength{\tabcolsep}{2pt}
\fontsize{8.5pt}{10pt}\selectfont
\begin{longtable}{lcccccccccc}
\caption{Operating characteristics for the binary endpoint.}\label{tab3}\\
\toprule
\textbf{Sc.} & \textbf{Design} & \textbf{max $N$} & \textbf{$\mathrm{FWER}_{00}$} & \textbf{$P(S)$} & \textbf{$\max_{\mathrm{grid}}\mathrm{FWER}_{01}$} & \textbf{$\widehat{U}_S$} & \textbf{$\mathrm{POWER}_{01}$} & \textbf{$\mathrm{POWER}_{11}$} & \textbf{Disjunctive power} & \textbf{$\mathrm{ESS}_0$} \\
\midrule
\endfirsthead
\multicolumn{11}{l}{\small\textit{Table~\thetable{} (continued)}}\\
\toprule
\textbf{Sc.} & \textbf{Design} & \textbf{max $N$} & \textbf{$\mathrm{FWER}_{00}$} & \textbf{$P(S)$} & \textbf{$\max_{\mathrm{grid}}\mathrm{FWER}_{01}$} & \textbf{$\widehat{U}_S$} & \textbf{$\mathrm{POWER}_{01}$} & \textbf{$\mathrm{POWER}_{11}$} & \textbf{Disjunctive power} & \textbf{$\mathrm{ESS}_0$} \\
\midrule
\endhead
\endfoot
\bottomrule
\multicolumn{11}{@{}p{0.8\textwidth}@{}}{\setstretch{1}\footnotesize\raggedright\textit{Note:} {Metrics are defined in Section~\ref{sec2.4}. $P(S)$ is the exact NG-survival bound; $\widehat{U}_S = \max\{\widehat{\mathrm{FWER}}_{00}, P(S)\}$ is a plug-in estimate, with confidence-bound verification in Table~\ref{tabS12}. Bold values identify the larger profiled error component. BOOST results are exact; other results use $10^5$ simulations. BOOST retains its published sample sizes; BOP2-ENR-nE uses the subgroup quotas of BOP2-ENR, and BOP2-P the same maximum total sample size and planned mixture weight, drawing all patients from the mixture (Section~\ref{sec3.2}). BOP2-P errors use a joint-subgroup interpretation of its mixture claim (Section~\ref{sec3.2}). $\mathrm{POWER}_{11}$ is reported only for BOP2-ENR. $^{\dagger}$Published strong-control designs re-evaluated exactly; S1--S4 and S8 exceed 0.05 in $\mathrm{FWER}_{00}$. Locally re-optimized designs appear in Table~\ref{tabS11}.}}\\
\endlastfoot
S1 & BOOST (weak, publ.) & 124 & \textbf{0.0713} & 0.0240 & 0.0240 & 0.0713 & 0.802 & --- & 0.813 & 39.9 \\
 & BOOST (strong, publ.)$^{\dagger}$ & 126 & \textbf{0.0753} & 0.0240 & 0.0240 & 0.0753 & 0.825 & --- & 0.836 & 41.2 \\
 & BOP2-P & 120 & 0.0258 & 1.0000 & \textbf{1.0000} & 1.0000 & 0.568 & --- & 0.954 & 68.5 \\
 & BOP2-ENR-nE & 120 & 0.0691 & 1.0000 & \textbf{1.0000} & 1.0000 & 0.874 & --- & 0.956 & 87.4 \\
 & BOP2-ENR & 120 & 0.0404 & 0.0426 & \textbf{0.0432} & 0.0426 & 0.859 & 0.540 & 0.916 & 52.6 \\
S2 & BOOST (weak, publ.) & 38 & 0.0370 & 0.2262 & \textbf{0.2262} & 0.2262 & 0.696 & --- & 0.798 & 15.4 \\
 & BOOST (strong, publ.)$^{\dagger}$ & 44 & \textbf{0.0533} & 0.0038 & 0.0038 & 0.0533 & 0.807 & --- & 0.816 & 17.6 \\
 & BOP2-P & 40 & 0.0352 & 1.0000 & \textbf{0.9994} & 1.0000 & 0.622 & --- & 0.887 & 18.3 \\
 & BOP2-ENR-nE & 40 & 0.0503 & 1.0000 & \textbf{1.0000} & 1.0000 & 0.813 & --- & 0.959 & 35.4 \\
 & BOP2-ENR & 40 & \textbf{0.0240} & 0.0237 & 0.0239 & 0.0240 & 0.778 & 0.454 & 0.873 & 24.2 \\
S3 & BOOST (weak, publ.) & 59 & 0.0271 & 0.0342 & \textbf{0.0342} & 0.0342 & 0.711 & --- & 0.755 & 22.1 \\
 & BOOST (strong, publ.)$^{\dagger}$ & 60 & \textbf{0.0725} & 0.0342 & 0.0342 & 0.0725 & 0.823 & --- & 0.847 & 23.3 \\
 & BOP2-P & 60 & 0.0291 & 1.0000 & \textbf{0.9941} & 1.0000 & 0.494 & --- & 0.935 & 32.4 \\
 & BOP2-ENR-nE & 60 & 0.0629 & 1.0000 & \textbf{1.0000} & 1.0000 & 0.819 & --- & 0.876 & 31.7 \\
 & BOP2-ENR & 60 & 0.0435 & 0.0450 & \textbf{0.0456} & 0.0450 & 0.809 & 0.555 & 0.853 & 22.2 \\
S4 & BOOST (weak, publ.) & 52 & 0.0495 & 0.0856 & \textbf{0.0856} & 0.0856 & 0.732 & --- & 0.802 & 18.6 \\
 & BOOST (strong, publ.)$^{\dagger}$ & 57 & \textbf{0.0581} & 0.0504 & 0.0504 & 0.0581 & 0.843 & --- & 0.864 & 21.9 \\
 & BOP2-P & 60 & 0.0385 & 1.0000 & \textbf{0.9885} & 1.0000 & 0.518 & --- & 0.975 & 35.9 \\
 & BOP2-ENR-nE & 60 & 0.0503 & 1.0000 & \textbf{1.0000} & 1.0000 & 0.864 & --- & 0.949 & 37.7 \\
 & BOP2-ENR & 60 & 0.0378 & 0.0451 & \textbf{0.0458} & 0.0451 & 0.855 & 0.687 & 0.926 & 25.7 \\
S5 & BOOST (weak, publ.) & 57 & 0.0623 & 0.1426 & \textbf{0.1426} & 0.1426 & 0.781 & --- & 0.802 & 25.5 \\
 & BOOST (strong, publ.)$^{\dagger}$ & 58 & \textbf{0.0282} & 0.0226 & 0.0226 & 0.0282 & 0.722 & --- & 0.769 & 30.1 \\
 & BOP2-P & 60 & 0.0249 & 1.0000 & \textbf{1.0000} & 1.0000 & 0.659 & --- & 0.903 & 36.4 \\
 & BOP2-ENR-nE & 60 & 0.0617 & 1.0000 & \textbf{1.0000} & 1.0000 & 0.814 & --- & 0.913 & 43.6 \\
 & BOP2-ENR & 60 & \textbf{0.0432} & 0.0237 & 0.0239 & 0.0432 & 0.794 & 0.431 & 0.857 & 35.1 \\
S6 & BOOST (weak, publ.) & 89 & 0.0508 & 0.1426 & \textbf{0.1426} & 0.1426 & 0.771 & --- & 0.785 & 35.9 \\
 & BOOST (strong, publ.)$^{\dagger}$ & 96 & \textbf{0.0372} & 0.0038 & 0.0038 & 0.0372 & 0.768 & --- & 0.777 & 40.5 \\
 & BOP2-P & 80 & 0.0266 & 1.0000 & \textbf{1.0000} & 1.0000 & 0.705 & --- & 0.891 & 47.4 \\
 & BOP2-ENR-nE & 80 & 0.0557 & 1.0000 & \textbf{1.0000} & 1.0000 & 0.840 & --- & 0.935 & 62.6 \\
 & BOP2-ENR & 80 & \textbf{0.0446} & 0.0237 & 0.0239 & 0.0446 & 0.824 & 0.443 & 0.887 & 53.2 \\
\pagebreak
S7 & BOOST (weak, publ.) & 36 & 0.0416 & 0.1855 & \textbf{0.1855} & 0.1855 & 0.738 & --- & 0.823 & 19.1 \\
 & BOP2-P & 40 & 0.0245 & 1.0000 & \textbf{0.9994} & 1.0000 & 0.522 & --- & 0.918 & 21.9 \\
 & BOP2-ENR-nE & 40 & 0.0630 & 1.0000 & \textbf{1.0000} & 1.0000 & 0.709 & --- & 0.823 & 22.3 \\
 & BOP2-ENR & 40 & \textbf{0.0410} & 0.0237 & 0.0239 & 0.0410 & 0.684 & 0.514 & 0.774 & 16.8 \\
S8 & BOOST (weak, publ.) & 42 & 0.0499 & 0.0815 & \textbf{0.0815} & 0.0815 & 0.743 & --- & 0.808 & 17.3 \\
 & BOOST (strong, publ.)$^{\dagger}$ & 54 & \textbf{0.0619} & 0.0257 & 0.0257 & 0.0619 & 0.834 & --- & 0.856 & 21.4 \\
 & BOP2-P & 40 & 0.0261 & 1.0000 & \textbf{0.9946} & 1.0000 & 0.508 & --- & 0.887 & 21.8 \\
 & BOP2-ENR-nE & 40 & 0.0459 & 1.0000 & \textbf{1.0000} & 1.0000 & 0.766 & --- & 0.940 & 37.9 \\
 & BOP2-ENR & 40 & 0.0388 & 0.0463 & \textbf{0.0466} & 0.0463 & 0.755 & 0.620 & 0.896 & 27.2 \\
\end{longtable}
\end{center}
\newpage
\begin{table}[htbp]
\centering
\caption{Summary of BOP2-ENR operating characteristics by endpoint.}\label{tabsum}
{\fontsize{8.5pt}{10pt}\selectfont
\begin{tabular}{lccccccc}
\toprule
\textbf{Endpoint} & \textbf{$P(S)$ exact} & \textbf{$\widehat{U}_S$} & \textbf{$\mathrm{FWER}_S$} & \textbf{$\mathrm{POWER}_{01}$} & \textbf{$\mathrm{POWER}_{11}$} & \textbf{Disjunctive power} & \textbf{$\mathrm{ESS}_0$} \\
\midrule
Binary & 0.024--0.046 & 0.024--0.046 & 0.024--0.047 & 0.68--0.86 (4) & 0.43--0.69 & 0.77--0.93 & 16.8--53.2 \\
Co-primary & 0.028--0.049 & 0.031--0.049 & 0.031--0.050 & 0.76--0.94 (7) & 0.39--0.74 & 0.89--0.98 & 19.7--53.1 \\
Efficacy-toxicity & 0.034--0.046 & 0.036--0.048 & 0.036--0.048 & 0.51--0.81 (1) & 0.20--0.48 & 0.57--0.85 & 22.0--56.2 \\
\bottomrule
\end{tabular}}
\tabnote{Entries are ranges across the eight scenarios; parentheses give the number with $\mathrm{POWER}_{01} \ge 0.80$. Metrics are defined in Section~\ref{sec2.4}. For efficacy-toxicity, $\mathrm{ESS}_0$ and $\mathrm{POWER}_{01}$ use the efficacy-null/tolerable-toxicity configuration. Full comparisons appear in Tables~\ref{tab3}, \ref{tab6} and \ref{tab9}.}
\end{table}

\clearpage
%% ===================== Supporting Information =====================
\addtocontents{toc}{\protect\setcounter{tocdepth}{2}}

\singlespacing
\setcounter{page}{1}
\setcounter{table}{0}
\renewcommand{\thetable}{S\arabic{table}}
\setcounter{figure}{0}
\renewcommand{\thefigure}{S\arabic{figure}}

\begin{center}
{\large Supporting Information for}\\[4pt]
{\large\bfseries BOP2-ENR: a Bayesian optimal phase II design for adaptive enrichment\\ with family-wise error rate control and a bound on joint-claim power}\\[8pt]
Kentaro Takeda, Belay B. Yimer, Masahiro Kojima
\end{center}

\begin{spacing}{0.92}\small\tableofcontents\end{spacing}

\SIsec{Appendix S1: Proofs, attainment of the bound, and empirical tightness}

{\textbf{Proof of the Lemma (least-favorable point of the global null).} Every decision in the trial, futility at each interim look and each claim at the final look, is a monotone function of the response counts: raising an NG count can only prevent an NG futility stop and can only help the pooled claim, and raising a PG count can only prevent a PG stop and help either claim. The binomial family is stochastically increasing in $p$, and the looks are nested, so the vector of counts at all looks is stochastically increasing in $p^-$ and in $p^+$ under the usual coupling; the any-claim event is an increasing event of that vector, so its probability is non-decreasing in each rate. $\square$

For the multi-criterion endpoints the same argument applies to each aggregated count under independence, with the toxicity count entering with the opposite sign: every decision probability is non-decreasing in each efficacy rate and non-increasing in the toxicity rate. For co-primary endpoints, whose null is the intersection of the two criteria, the least-favorable point is the null vector with both criteria at threshold. For efficacy-toxicity the least-favorable point of the efficacy-null face is $(\theta_E, \theta_T) = (\phi_{01}, 0)$, inefficacious and perfectly safe, and that of the toxicity-null face is $(1, \phi_{02})$, perfectly efficacious and at the toxicity threshold, or $(\phi_{01}^+, \phi_{02})$ where the efficacy ordering caps the NG efficacy at the PG null; at the first two one margin is degenerate (toxicity zero, or efficacy one), so the joint distribution is determined by the other margin and is unaffected by any association between the two outcomes; the third, and the co-primary null vector, are non-degenerate in both margins, and their bounds depend on the association. $\mathrm{FWER}_{00}$ is therefore evaluated at these points throughout.

\textbf{Remark (attainment of the bound and invariance to $\lambda_E$).} \textit{For the single binary endpoint, let $n_{R-1}^-$ be the NG cohort size at the last interim look, $b_F^-(n_{R-1}^-)$ the corresponding integer futility boundary (Section~\ref{sec2.6}), $b_E^{\mathrm{pool}}$ the pooled-claim boundary for the combined cohort of $N^- + N^+$ patients, and $m_{\min} = \min\{x_N^- : \text{an NG trajectory with final count } x_N^- \text{ survives every look}\}$ the smallest final NG response count compatible with $S$, so that $m_{\min} \ge b_F^-(n_{R-1}^-) + 1$, with equality when the futility boundaries are non-decreasing in the look index, as for the designs reported here. Then, along the profiling path, $\mathrm{FWER}_{01}(\psi) \to P(S)$ as $\psi \to 1$ if and only if}
\begin{equation*}
b_E^{\mathrm{pool}} \;\le\; N^+ + m_{\min}.
\end{equation*}
\textit{Proof.} As $\psi \to 1$ every PG patient responds and PG survives its own futility looks with probability tending to one, so the joint claim occurs if and only if $S$ occurs and the pooled count $N^+ + x^-$ reaches $b_E^{\mathrm{pool}}$, where $x^-$ is the final NG response count. On $S$ the final count satisfies $x^- \ge m_{\min}$ by definition, so the claim is certain on $S$ under the displayed condition and $\mathrm{FWER}_{01}(\psi) \to P(S)$. If the condition fails, the event $S \cap \{x^- = m_{\min}\}$ has positive probability (a surviving trajectory attaining $m_{\min}$ exists by definition, and $P(S) > 0$) and yields no claim, so the limit is strictly below $P(S)$. $\square$

The condition is monotone in $\lambda_E$: $b_E^{\mathrm{pool}}$ is non-decreasing in $\lambda_E$, so once the condition holds it continues to hold for every smaller $\lambda_E$, and over that range the limiting mixed-configuration error as PG efficacy tends to one equals $P(S)$ and is therefore invariant to $\lambda_E$; the maximum evaluated on the finite profiling grid ($\psi = 0.99$ for the binary endpoint, $(0.97, 0.97)$ for co-primary endpoints) is a numerical approximation to this supremum. For the multi-criterion endpoints the condition becomes that every NG terminal state reachable on $S$ lies, once combined with the extreme PG alternative, in the pooled claim region defined by $\mathrm{op}_E$; for co-primary endpoints, where the extreme PG alternative satisfies both criteria and the disjunctive rule is monotone in each count, the condition was verified for each of the eight reported designs (it is not automatic for an arbitrary $\lambda_E$), whereas it fails for efficacy-toxicity, where the toxicity criterion is worsened rather than helped by NG events, so that attainment there is a matter of degree rather than of a single integer inequality.

\textbf{Scope of the Remark.} The tightness of the bound in every scenario-by-endpoint combination and its insensitivity to $\lambda_E$ are reported in Section~\ref{sec4.0} and Tables~\ref{tabS12} and \ref{tabS13}. Equality is not asserted in general, for the reasons given in the Remark: NG outcomes remain random even as PG becomes fully effective, the pooled posterior depends on the combined denominator as well as the combined counts, and for efficacy-toxicity the toxicity criterion is worsened by NG events. Accordingly, the near-invariance of the profiled maximum to $\lambda_E$ and the statement that the NG gate is the binding constraint on the mixed configuration are regularities of the settings studied rather than consequences of Proposition (a). Proposition (a) concerns the NG-null/PG-active direction only; the reverse direction, excluded by the monotone-activity assumption, is examined separately in the Discussion.

\textbf{The joint-power ceiling.} In the settings studied the gap between $\mathrm{POWER}_{11}$ and the relevant ceiling ranged from 0.007 to 0.104 for the binary endpoint, from 0.000 to 0.023 for co-primary endpoints, and from 0.001 to 0.043 for efficacy-toxicity monitoring (Table~\ref{tabS14}).}

\SIsec{Appendix S2: Number of interim analyses}

{This appendix reports the operating characteristics of BOP2-ENR recalibrated from scratch at $R = 2, \dots, 6$ total looks (equivalently $R - 1$ interim futility analyses plus a final analysis) for each of the eight scenarios and each endpoint; $R = 4$ is used in the main manuscript. Tables~\ref{tabS1}, \ref{tabS4} and \ref{tabS7} report the strong FWER, power and expected sample size averaged over the scenarios for the binary, co-primary and efficacy-toxicity endpoints respectively (means are over the eight scenarios, except for co-primary endpoints at $R = 2$, where only five scenarios were feasible; the co-primary means at $R = 2$ and at larger $R$ therefore summarize different scenario sets and are not directly comparable); the distribution of the look at which enrichment occurred under the retained schedule is given in Tables~\ref{tabS3}, \ref{tabS6} and \ref{tabS9}.

Two features are common to all three endpoints. At $R = 2$ (a single interim analysis, the schedule of the two-stage designs), with the prespecified sample sizes, tuning grid and bound-based calibration criterion, no gate with exact bound at most $\alpha$ was found in three of the eight co-primary scenarios, so that no feasible candidate exists there under that criterion, while every schedule with $R \ge 3$ is feasible and satisfied the exact mixed-bound criterion and the Monte Carlo global-null calibration criterion in all eight for every endpoint; the profiled diagnostic exceeded 0.05 by at most 0.0006 in three of the 120 schedule-by-scenario evaluations (co-primary S1 at $R = 4$, binary S3 and efficacy-toxicity S3 at $R = 6$), where the exact bound is 0.049 and the excess is Monte Carlo noise at $6 \times 10^4$ replicates. Mean expected sample size falls substantially but not monotonically with the number of looks: for the binary and efficacy-toxicity endpoints by 6.4 to 7.0 patients from $R = 2$ to $R = 4$ and by a further 2.1 to 2.5 patients from $R = 4$ to $R = 6$, and for co-primary endpoints by 2.9 patients from $R = 4$ to $R = 6$.

$R = 4$ captures most of the efficiency available from interim monitoring. $R = 2$ is feasible for the binary and efficacy-toxicity endpoints at a cost of about seven patients in expected sample size, and for co-primary endpoints in five of eight scenarios under the prespecified calibration criterion.}

\SIsec{Appendix S3: Verification of the computational engines and reproducibility of the published two-stage operating characteristics}

\textbf{Exact computation of the bounds.} The bounds $P(S)$ and $P(S^{\ast})$ are computed exactly rather than by simulation. The futility decision at each look depends on the aggregated counts of the criteria only, $a = x_1 + x_2$ and, for $K = 4$, $b = x_1 + x_3$; increments between looks are multinomial and independent of the past, so $(a, b)$ is a Markov chain on a grid of at most $(n+1)^2$ states. Its distribution is propagated look by look, the states that trip the futility rule (under $\mathrm{op}_F$) being set to zero at each interim look, and the surviving mass at the last interim look is the bound. For $K = 2$ the chain is univariate. Because $\theta$ enters only through the increment distribution, any four-category vector may be used, which is how the association sensitivity of Table~\ref{tabS15} is obtained. The recursion agrees with the Monte Carlo estimator to within Monte Carlo error in every scenario, and the selected designs were additionally re-evaluated from an independent seed at $4 \times 10^5$ replicates (Table~\ref{tabS12}). The integer boundaries of Tables~\ref{tab2}, \ref{tab5} and \ref{tab8}, including the subgroup-specific safety boundaries, are generated from the same routine that the verification reads, and the verification compares them with the boundary-crossing and claim indicators recorded by the simulator itself over $2 \times 10^4$ trials per scenario under the global alternative, so that the enrichment path and the continue-both path are both exercised.

The FWER machinery was verified in the $K = 2$ case against exact enumeration for a two-stage benchmark, using binomial and two-probability Poisson-binomial convolutions. Monte Carlo and exact evaluation agreed within Monte Carlo error throughout: with $2 \times 10^6$ replicates the largest discrepancy across the eight scenarios was 0.0003 in the weak FWER, 0.0006 in disjunctive power and 0.02 patients in expected sample size, against a Monte Carlo standard error of about 0.0002 at an error rate of 0.05. For the scenario with $\phi_0 = 0.03$ the weak FWER was 0.0711 by simulation against 0.0713 exactly, disjunctive power 0.8130 against 0.8127, and expected sample size 39.85 against 39.85; for $\phi_0 = 0.10$ the corresponding values were 0.0270 against 0.0271, 0.7554 against 0.7551, and 22.14 against 22.14. Both evaluations apply the second-stage cut-offs as strict inequalities, the convention discussed below. The binary endpoint is handled by the same Dirichlet-multinomial engine as the multi-category ones, at $K = 2$, so no separate binary implementation exists to disagree with it. What is verified instead is that the engine's futility boundary at $K = 2$ coincides with the beta-binomial boundary computed directly, and that the boundaries tabulated from the posterior rules reproduce the posterior decisions exactly in all 24 scenario-by-endpoint combinations; both checks are part of the regression suite.

\textbf{Conventions for the exact recomputation of the published two-stage designs.} (i) Expected sample sizes are computed under the routing rule of Section~\ref{sec2.3}, which excludes the NG-only continuation path; the published values appear to follow an additively separable expression that includes that path, and differ from the recomputation by up to about 3.0 patients (S2, weak variant), the recomputed values being the smaller. (ii) Second-stage cut-offs are applied as strict inequalities, the convention under which the published type I error rates are reproduced; matching the published power instead points to the adjacent boundary, for the weak- and the strong-control designs alike. Both conventions act at the level of integer-boundary rounding and do not affect $\mathrm{FWER}_{01}$, which is governed by the futility gate.

One published design is excluded outright: the S7 strong-control row lists stage sizes $N = (14, 15)$ against a stated total of 39, and no reading of the row reproduces its reported operating characteristics, so it is omitted from Table~\ref{tab3} pending clarification; no re-optimized counterpart is offered, since the published sample sizes do not determine a unique starting design; all comparisons involving the strong variant therefore cover seven scenarios. These are conventions of the recomputation rather than claims of error; they are stated because the published designs serve as comparators.

\SIsec{Appendix S4: Full calibration specification}

\begin{figure}[htbp]
\centering
\includegraphics[width=\textwidth]{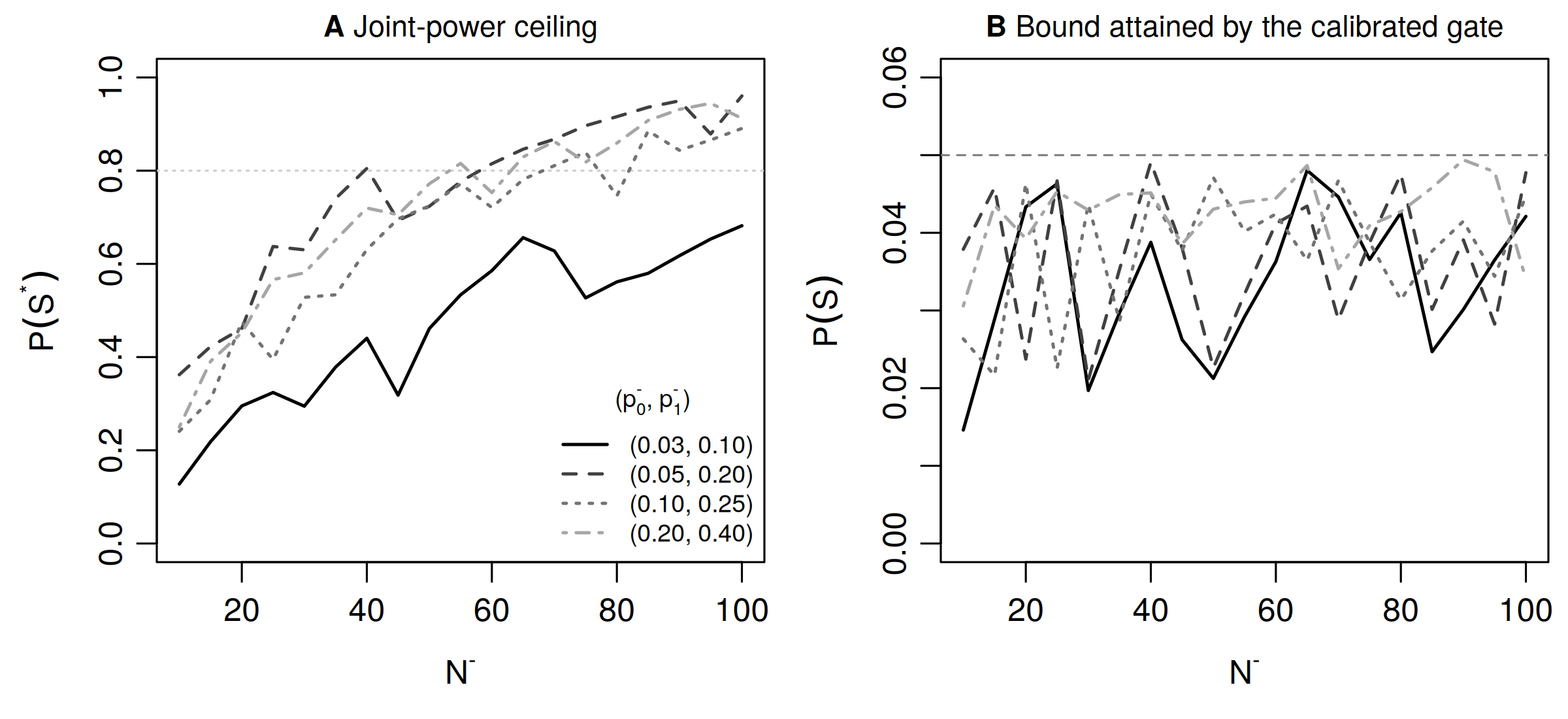}
\caption{NG sample size and the joint-power ceiling. Four equally spaced looks, with the NG gate calibrated separately at each $N^-$ using Appendix S4. \textbf{A} Exact joint-power ceilings $P(S^{\ast})$ for the NG response-rate pairs of S1--S4; the dotted line marks 0.80. \textbf{B} Exact NG-null survival bounds $P(S)$; the dashed line marks 0.05. Irregularities reflect discrete count boundaries.}\label{fig3}\addcontentsline{toc}{subsection}{Figure S1: NG sample size and the joint-power ceiling}
\end{figure}

Calibration used a deterministic seeded sequential search (seeds 11 calibration, 77 evaluation, 3 BOP2-P); candidates consume independent random streams, so this is reproducible but is not a common-random-numbers scheme. Recalibration of the binary endpoint from two further seeds (22, 33) reproduced the selected tuning parameters exactly or at an adjacent grid value in every scenario, leaving the reported operating characteristics unchanged at the reported precision. Step 1 scanned $\gamma_- \in \{0.35, 0.20, 0.10, 0.05\}$ (in that order) and, within each, $\lambda_- \in \{0.50, 0.70, 0.85, 0.92, 0.95, 0.97, 0.985, 0.995, 0.999\}$, accepting the first pair whose exact bound $P(S)$ at the least-favorable NG null is at most $\alpha$ (no Monte Carlo margin; for efficacy-toxicity, Step 1b then fixed $\lambda_T^{\pm}$ as the smallest values with exact size at most $\alpha$ on the grid 0.80, 0.85, 0.90, 0.92, 0.94, 0.95, 0.96, 0.97, 0.975, 0.98, 0.985, 0.99, 0.995); ties are resolved by taking the first feasible gate in the prespecified scan order. Step 2 scanned $\lambda_+ \in \{0.60, 0.35, 0.20, 0.10, 0.05, 0.02\}$ with $\gamma_+ = 0.50$ fixed. Step 3 chose the smallest $\lambda_E$ on the grid $\{0.80, 0.85, 0.88, 0.90, 0.92, 0.94, 0.95, 0.96, 0.97, 0.975, 0.98, 0.985, 0.99, 0.995, 0.998\}$ with $\mathrm{FWER}_{00} \le 0.9\alpha$ at $3 \times 10^4$ replicates, $\mathrm{FWER}_{00}$ being the maximum over the admissible null corners of Section~\ref{sec2.4}. Among candidates attaining $\mathrm{POWER}_{01} \ge 0.80$ at the design alternative the one with the smallest $\mathrm{ESS}_0$ was retained; if none attained it, the candidate with the largest power was retained. The $0.9\alpha$ margin applies to this Monte Carlo constraint only, as the admissibility allowance for Monte Carlo error at the calibration sample size; retained parameters were then evaluated at $10^5$ replicates. Figure~\ref{fig3} shows the Step 1 output, the calibrated bound $P(S)$ and the joint-power ceiling $P(S^{\ast})$, as a function of $N^-$ for the binary scenarios S1--S4.

\clearpage
\SIsec{Supporting tables: single binary endpoint}

\textit{Common conventions.} NG and PG denote the biomarker-negative and biomarker-positive cohorts. Error and power measures follow Section~\ref{sec2.4} of the main article. Unless stated otherwise, designs use four equally spaced looks (three interim analyses and a final analysis) with $\alpha = 0.05$. For efficacy-toxicity, the reported certifying bound is the maximum over the two NG-null faces, and $\mathrm{ESS}_0$ and $\mathrm{POWER}_{01}$ use the efficacy-null/tolerable-toxicity configuration.

\begin{table}[H]
\centering
\caption{Sensitivity to the number of looks: binary endpoint.}\label{tabS1}\addcontentsline{toc}{subsection}{Table S1: Sensitivity to the number of looks: binary endpoint}
{\fontsize{9pt}{10.8pt}\selectfont
\begin{tabular}{lccccc}
\toprule
\shortstack{\textbf{Total}\\\textbf{looks $R$}} & \shortstack{\textbf{Interim}\\\textbf{futility looks}} & \shortstack{\textbf{Scenarios with profiled}\\\textbf{diagnostic $\mathrm{FWER}_S \le \alpha$}} & \shortstack{\textbf{Mean}\\\textbf{$\mathrm{POWER}_{01}$}} & \shortstack{\textbf{Mean}\\\textbf{disjunctive power}} & \shortstack{\textbf{Mean}\\\textbf{$\mathrm{ESS}_0$}} \\
\midrule
2 & 1 & 8/8 & 0.813 & 0.879 & 41.0 \\
3 & 2 & 8/8 & 0.799 & 0.877 & 33.0 \\
4 & 3 & 8/8 & 0.808 & 0.889 & 34.0 \\
5 & 4 & 8/8 & 0.791 & 0.865 & 31.0 \\
6 & 5 & 7/8 & 0.788 & 0.872 & 31.5 \\
\bottomrule
\end{tabular}}
\tabnote{Each scenario was recalibrated for each total number of looks $R$, including the final analysis. Power and sample-size entries are scenario averages. The count column summarizes the profiled diagnostic, not calibration feasibility; an estimate may exceed 0.05 despite an exact bound below 0.05 (Appendix S2). Operating characteristics use $6 \times 10^4$ simulations per evaluated design.}
\end{table}

\begin{table}[H]
\centering
\caption{Calibrated parameters for the binary endpoint.}\label{tabS2}\addcontentsline{toc}{subsection}{Table S2: Calibrated parameters for the binary endpoint}
{\fontsize{9pt}{10.8pt}\selectfont
\begin{tabular}{lcccc}
\toprule
\textbf{Sc.} & \textbf{NG futility $\lambda_-$} & \textbf{NG futility $\gamma_-$} & \textbf{PG futility $\lambda_+$} & \textbf{Efficacy cutoff $\lambda_E$} \\
\midrule
S1 & 0.920 & 0.35 & 0.02 & 0.900 \\
S2 & 0.950 & 0.35 & 0.02 & 0.940 \\
S3 & 0.950 & 0.35 & 0.35 & 0.880 \\
S4 & 0.985 & 0.35 & 0.60 & 0.940 \\
S5 & 0.950 & 0.35 & 0.10 & 0.920 \\
S6 & 0.950 & 0.35 & 0.35 & 0.950 \\
S7 & 0.950 & 0.35 & 0.20 & 0.880 \\
S8 & 0.950 & 0.35 & 0.05 & 0.950 \\
\bottomrule
\end{tabular}}
\tabnote{Futility cutoffs are $C_F^{\pm}(n) = \lambda_{\pm}(n/N^{\pm})^{\gamma_{\pm}}$, with $\gamma_+ = 0.50$. The common final efficacy cutoff is $\lambda_E$.}
\end{table}

The calibrated cutoffs show the division of labor set out in Sections~\ref{sec2.5} and \ref{sec2.5b}: the NG futility cutoff $\lambda_-$ is uniformly aggressive because it bounds the mixed-configuration component of the strong-control error rate, whereas the PG cutoff $\lambda_+$ is far more permissive because it serves efficiency and an aggressive value would forfeit power.

\begin{center}
\fontsize{9pt}{10.8pt}\selectfont
\begin{longtable}{lcccccc}
\caption{Enrichment timing and trial stopping: binary endpoint.}\label{tabS3}\addcontentsline{toc}{subsection}{Table S3: Enrichment timing and trial stopping: binary endpoint}\\
\toprule
\textbf{Sc.} & \textbf{Configuration} & \textbf{$L_1$} & \textbf{$L_2$} & \textbf{$L_3$} & \textbf{Never enriched} & \textbf{Trial stopped (PG futility)} \\
\midrule
\endfirsthead
\multicolumn{7}{l}{\small\textit{Table~\thetable{} (continued)}}\\
\toprule
\textbf{Sc.} & \textbf{Configuration} & \textbf{$L_1$} & \textbf{$L_2$} & \textbf{$L_3$} & \textbf{Never enriched} & \textbf{Trial stopped (PG futility)} \\
\midrule
\endhead
\endfoot
\bottomrule
\multicolumn{7}{@{}p{0.85\textwidth}@{}}{\footnotesize\raggedright\textit{Note:} {Entries are percentages of $10^5$ simulated trials. $L_r$ denotes first NG freezing at interim look $r$; these categories and ``Never enriched'' sum to 100\%, apart from rounding. PG stopping is reported separately and is not an additional mutually exclusive category. ``Never enriched'' includes trials stopped before enrichment and is not the NG-survival bound.}}\\
\endlastfoot
S1 & global null & 88.0 & 2.5 & 1.1 & 8.4 & 54.4 \\
 & mixed & 88.0 & 5.2 & 2.2 & 4.6 & 3.9 \\
 & global alt & 39.5 & 3.4 & 1.2 & 55.9 & 3.9 \\
S2 & global null & 77.5 & 15.7 & 2.4 & 4.5 & 46.1 \\
 & mixed & 77.5 & 15.7 & 4.3 & 2.5 & 1.4 \\
 & global alt & 33.0 & 13.6 & 7.6 & 45.8 & 1.4 \\
S3 & global null & 29.9 & 6.6 & 1.6 & 61.9 & 70.5 \\
 & mixed & 64.9 & 14.3 & 4.9 & 15.9 & 12.3 \\
 & global alt & 21.7 & 7.4 & 3.8 & 67.1 & 12.3 \\
S4 & global null & 45.3 & 9.2 & 3.2 & 42.3 & 69.6 \\
 & mixed & 65.8 & 16.5 & 9.9 & 7.8 & 4.7 \\
 & global alt & 16.4 & 4.7 & 6.3 & 72.6 & 4.7 \\
S5 & global null & 50.1 & 10.2 & 2.6 & 37.1 & 40.0 \\
 & mixed & 73.0 & 14.8 & 4.1 & 8.1 & 5.8 \\
 & global alt & 31.0 & 12.8 & 7.3 & 48.9 & 5.8 \\
S6 & global null & 64.3 & 10.9 & 2.7 & 22.1 & 39.7 \\
 & mixed & 76.4 & 15.4 & 4.3 & 4.0 & 2.1 \\
 & global alt & 32.5 & 13.3 & 7.6 & 46.6 & 2.1 \\
S7 & global null & 31.5 & 6.4 & 1.8 & 60.4 & 59.4 \\
 & mixed & 64.3 & 13.1 & 3.6 & 19.1 & 17.0 \\
 & global alt & 19.8 & 7.8 & 4.4 & 68.0 & 17.0 \\
S8 & global null & 59.3 & 17.2 & 15.1 & 8.4 & 10.7 \\
 & mixed & 59.3 & 19.2 & 16.8 & 4.7 & 0.3 \\
 & global alt & 17.0 & 6.1 & 13.8 & 63.1 & 0.3 \\
\end{longtable}
\end{center}

Enrichment generally occurred at the first two interim analyses (Tables~\ref{tabS3}, \ref{tabS6} and \ref{tabS9}): under the mixed configuration the NG cohort was frozen at $L_1$ in 59--88\% (binary), 74--92\% (co-primary) and 63--88\% (efficacy-toxicity) of replicates, a direct consequence of the aggressive NG gate that the bound-based calibration requires; the co-primary profile is the most front-loaded because its conjunctive futility rule is compensated through a smaller $\gamma_-$, which raises the cutoff at the early looks. Under the global alternative the NG cohort was retained in 46--73\%, 40--74\% and 36--65\% of replicates respectively, so the aggressive gate does not simply discard a genuinely active biomarker-negative subgroup. Because most of the enrichment decision is made early, later looks add little for the NG cohort, and the residual benefit of longer schedules accrues mainly through PG futility stopping.

``Never enriched'' includes trials stopped for PG futility before enrichment and therefore differs from the counterfactual NG-survival probability $P(S)$. With $E$ the enrichment event, $G$ the event that the PG cohort passes every interim look and $H$ the event of a PG-futility stop before the NG cohort is frozen (a PG stop taking precedence at a common look), $\bar{E} = (S \cap G) \,\dot\cup\, H$, so that $P(\bar{E}) = P(S)\,P(G) + P(H)$ under independent subgroup streams, whereas the PG-stop column also counts stops after enrichment. The never-enriched column is thus an upper envelope on $P(S)$ inflated by PG-futility stops (binary S3: 15.9\% never enriched, 12.3\% PG stops, exact $P(S) = 4.5$\%; efficacy-toxicity S3: 16.0\%, 13.0\%, 4.5\%).

\SIsec{Supporting tables: co-primary endpoints}

\begin{table}[H]
\centering
\caption{Scenario parameters and subgroup sample sizes, co-primary endpoints.}\label{tabScp}\addcontentsline{toc}{subsection}{Table S4: Scenario parameters and subgroup sample sizes, co-primary endpoints}
{\fontsize{8.5pt}{10pt}\selectfont\setlength{\tabcolsep}{4pt}
\begin{tabular}{lcccccccc}
\toprule
\textbf{Sc.} & \textbf{Null $(p_0^-,p_0^+)$} & \textbf{Alt. $(p_1^-,p_1^+)$} & \textbf{PFS6 null $(q_0^-,q_0^+)$} & \textbf{PFS6 alt. $(q_1^-,q_1^+)$} & \textbf{$N^-$} & \textbf{$N^+$} & \textbf{Pooled null $\phi_0^{\mathrm{pool}}$} & \textbf{Pooled PFS6 null} \\
\midrule
S1 & (0.03, 0.03) & (0.10, 0.15) & (0.18, 0.18) & (0.25, 0.30) & 80 & 40 & 0.030 & 0.180 \\
S2 & (0.05, 0.05) & (0.20, 0.25) & (0.20, 0.20) & (0.35, 0.40) & 20 & 20 & 0.050 & 0.200 \\
S3 & (0.10, 0.10) & (0.25, 0.35) & (0.25, 0.25) & (0.40, 0.50) & 40 & 20 & 0.100 & 0.250 \\
S4 & (0.20, 0.20) & (0.40, 0.50) & (0.35, 0.35) & (0.55, 0.65) & 40 & 20 & 0.200 & 0.350 \\
S5 & (0.05, 0.10) & (0.20, 0.25) & (0.20, 0.25) & (0.35, 0.40) & 20 & 40 & 0.083 & 0.233 \\
S6 & (0.05, 0.20) & (0.20, 0.35) & (0.20, 0.35) & (0.35, 0.50) & 20 & 60 & 0.163 & 0.313 \\
S7 & (0.05, 0.10) & (0.25, 0.30) & (0.20, 0.25) & (0.40, 0.45) & 20 & 20 & 0.075 & 0.225 \\
S8 & (0.10, 0.20) & (0.30, 0.45) & (0.25, 0.35) & (0.45, 0.60) & 20 & 20 & 0.150 & 0.300 \\
\bottomrule
\end{tabular}}
\tabnote{Response rates and sample sizes are those of Table~\ref{tab1}; PFS6 rates are the response rates plus 0.15 ($q = \min\{p + 0.15, 0.95\}$), so that $\theta_0^{\pm}$ and $\theta_1^{\pm}$ are the cross-classified cell probabilities of response and PFS6 under independence. Pooled null thresholds are allocation-weighted (Section~\ref{sec2.3}); the pooled PFS6 threshold equals $\phi_0^{\mathrm{pool}} + 0.15$.}
\end{table}

\begin{table}[H]
\centering
\caption{Sensitivity to the number of looks: co-primary endpoints.}\label{tabS4}\addcontentsline{toc}{subsection}{Table S5: Sensitivity to the number of looks: co-primary endpoints}
{\fontsize{9pt}{10.8pt}\selectfont
\begin{tabular}{lccccc}
\toprule
\shortstack{\textbf{Total}\\\textbf{looks $R$}} & \shortstack{\textbf{Interim}\\\textbf{futility looks}} & \shortstack{\textbf{Scenarios with profiled}\\\textbf{diagnostic $\mathrm{FWER}_S \le \alpha$}} & \shortstack{\textbf{Mean}\\\textbf{$\mathrm{POWER}_{01}$}} & \shortstack{\textbf{Mean}\\\textbf{disjunctive power}} & \shortstack{\textbf{Mean}\\\textbf{$\mathrm{ESS}_0$}} \\
\midrule
2 & 1 & 5/8 & 0.828 & 0.915 & 39.1 \\
3 & 2 & 8/8 & 0.865 & 0.935 & 35.9 \\
4 & 3 & 7/8 & 0.866 & 0.928 & 33.3 \\
5 & 4 & 8/8 & 0.858 & 0.935 & 31.9 \\
6 & 5 & 8/8 & 0.849 & 0.934 & 30.4 \\
\bottomrule
\end{tabular}}
\tabnote{Each scenario was recalibrated for each total number of looks $R$, including the final analysis. The count column summarizes the profiled diagnostic. With $R = 2$, no feasible candidate was found for three scenarios within the prespecified calibration grid, and the power and sample-size averages include only the five feasible scenarios; feasibility and Monte Carlo diagnostic exceedances are distinguished in Appendix S2. Operating characteristics use $6 \times 10^4$ simulations per evaluated design.}
\end{table}

\begin{table}[H]
\centering
\caption{Calibrated parameters for co-primary endpoints.}\label{tabS5}\addcontentsline{toc}{subsection}{Table S6: Calibrated parameters for co-primary endpoints}
{\fontsize{9pt}{10.8pt}\selectfont
\begin{tabular}{lcccc}
\toprule
\textbf{Sc.} & \textbf{NG futility $\lambda_-$} & \textbf{NG futility $\gamma_-$} & \textbf{PG futility $\lambda_+$} & \textbf{Efficacy cutoff $\lambda_E$} \\
\midrule
S1 & 0.995 & 0.20 & 0.60 & 0.970 \\
S2 & 0.920 & 0.10 & 0.60 & 0.975 \\
S3 & 0.950 & 0.10 & 0.60 & 0.970 \\
S4 & 0.985 & 0.10 & 0.60 & 0.975 \\
S5 & 0.920 & 0.10 & 0.60 & 0.970 \\
S6 & 0.920 & 0.10 & 0.60 & 0.985 \\
S7 & 0.920 & 0.10 & 0.02 & 0.970 \\
S8 & 0.950 & 0.20 & 0.60 & 0.975 \\
\bottomrule
\end{tabular}}
\tabnote{Futility cutoffs follow Table~\ref{tabS2}, with $\gamma_+ = 0.50$. $\lambda_E$ is the common final efficacy cutoff; futility requires both criteria to fail, whereas either criterion suffices for a claim.}
\end{table}

Relative to the binary endpoint the whole calibration is displaced upward: $\gamma_-$ is reduced to make the gate bite earlier, a consequence of the conjunctive futility rule, and $\lambda_E$ rises to absorb the type I inflation of the disjunctive claim rule (Section~\ref{sec4.2} of the main article).

\begin{center}
\fontsize{9pt}{10.8pt}\selectfont
\begin{longtable}{lcccccc}
\caption{Enrichment timing and trial stopping: co-primary endpoints.}\label{tabS6}\addcontentsline{toc}{subsection}{Table S7: Enrichment timing and trial stopping: co-primary endpoints}\\
\toprule
\textbf{Sc.} & \textbf{Configuration} & \textbf{$L_1$} & \textbf{$L_2$} & \textbf{$L_3$} & \textbf{Never enriched} & \textbf{Trial stopped (PG futility)} \\
\midrule
\endfirsthead
\multicolumn{7}{l}{\small\textit{Table~\thetable{} (continued)}}\\
\toprule
\textbf{Sc.} & \textbf{Configuration} & \textbf{$L_1$} & \textbf{$L_2$} & \textbf{$L_3$} & \textbf{Never enriched} & \textbf{Trial stopped (PG futility)} \\
\midrule
\endhead
\endfoot
\bottomrule
\endlastfoot
S1 & global null & 51.2 & 6.8 & 3.4 & 38.5 & 54.5 \\
 & mixed & 73.7 & 11.2 & 7.3 & 7.8 & 3.2 \\
 & global alt & 23.4 & 3.9 & 4.4 & 68.3 & 3.2 \\
S2 & global null & 68.6 & 1.0 & 1.4 & 29.0 & 59.4 \\
 & mixed & 90.3 & 1.5 & 3.2 & 4.9 & 2.4 \\
 & global alt & 55.3 & 0.8 & 2.7 & 41.2 & 2.4 \\
S3 & global null & 73.8 & 4.4 & 1.6 & 20.3 & 44.4 \\
 & mixed & 85.5 & 7.3 & 2.9 & 4.4 & 0.7 \\
 & global alt & 33.1 & 4.2 & 1.2 & 61.6 & 0.7 \\
S4 & global null & 68.4 & 8.9 & 3.2 & 19.4 & 42.9 \\
 & mixed & 79.5 & 12.0 & 5.5 & 3.0 & 0.2 \\
 & global alt & 18.9 & 3.8 & 3.3 & 74.0 & 0.2 \\
S5 & global null & 84.2 & 1.1 & 1.9 & 12.8 & 41.9 \\
 & mixed & 91.7 & 1.5 & 3.3 & 3.5 & 0.8 \\
 & global alt & 56.2 & 0.7 & 2.7 & 40.4 & 0.8 \\
S6 & global null & 79.0 & 1.2 & 1.9 & 17.9 & 42.7 \\
 & mixed & 91.6 & 1.5 & 3.3 & 3.5 & 0.5 \\
 & global alt & 56.2 & 0.7 & 2.7 & 40.4 & 0.5 \\
S7 & global null & 92.0 & 1.6 & 3.3 & 3.2 & 0.3 \\
 & mixed & 92.1 & 1.5 & 3.3 & 3.1 & 0.0 \\
 & global alt & 43.1 & 0.4 & 1.4 & 55.1 & 0.0 \\
S8 & global null & 70.8 & 6.9 & 2.1 & 20.3 & 43.0 \\
 & mixed & 82.0 & 9.1 & 3.7 & 5.2 & 0.6 \\
 & global alt & 31.1 & 5.1 & 2.1 & 61.6 & 0.6 \\
\end{longtable}
\tabnote{Percentages of $10^5$ simulated trials; timing categories follow Table~\ref{tabS3}. PG stopping is a separate, overlapping measure, and ``Never enriched'' is not the NG-survival bound.}
\end{center}

\begin{table}[htbp]
\centering
\caption{Decision boundaries for co-primary endpoints.}\label{tab5}\addcontentsline{toc}{subsection}{Table S8: Decision boundaries for co-primary endpoints}
{\fontsize{9pt}{10.8pt}\selectfont
\begin{tabular}{lccccc}
\toprule
\textbf{Sc.} & \textbf{Criterion} & \textbf{Futility, NG} & \textbf{Futility, PG} & \textbf{PG-only claim} & \textbf{Pooled claim} \\
\midrule
S1 & ORR & $\le$ 1, 2, 4 & $\le$ 0, 0, 1 & $\ge$ 5 & $\ge$ 8 \\
 & PFS6 & $\le$ 5, 10, 15 & $\le$ 1, 3, 5 & $\ge$ 13 & $\ge$ 31 \\
S2 & ORR & $\le$ 1, 1, 2 & $\le$ 0, 0, 1 & $\ge$ 4 & $\ge$ 6 \\
 & PFS6 & $\le$ 2, 3, 5 & $\le$ 0, 1, 3 & $\ge$ 9 & $\ge$ 14 \\
S3 & ORR & $\le$ 2, 4, 5 & $\le$ 0, 1, 1 & $\ge$ 6 & $\ge$ 12 \\
 & PFS6 & $\le$ 4, 7, 11 & $\le$ 0, 2, 3 & $\ge$ 10 & $\ge$ 22 \\
S4 & ORR & $\le$ 3, 6, 10 & $\le$ 0, 1, 3 & $\ge$ 9 & $\ge$ 19 \\
 & PFS6 & $\le$ 5, 10, 15 & $\le$ 1, 3, 5 & $\ge$ 12 & $\ge$ 29 \\
S5 & ORR & $\le$ 1, 1, 2 & $\le$ 0, 2, 3 & $\ge$ 9 & $\ge$ 10 \\
 & PFS6 & $\le$ 2, 3, 5 & $\le$ 1, 4, 7 & $\ge$ 16 & $\ge$ 21 \\
S6 & ORR & $\le$ 1, 1, 2 & $\le$ 2, 5, 9 & $\ge$ 20 & $\ge$ 21 \\
 & PFS6 & $\le$ 2, 3, 5 & $\le$ 4, 10, 16 & $\ge$ 30 & $\ge$ 35 \\
S7 & ORR & $\le$ 1, 1, 2 & $\le$ ---, ---, 0 & $\ge$ 6 & $\ge$ 7 \\
 & PFS6 & $\le$ 2, 3, 5 & $\le$ ---, 0, 0 & $\ge$ 10 & $\ge$ 15 \\
S8 & ORR & $\le$ 1, 2, 3 & $\le$ 0, 1, 3 & $\ge$ 9 & $\ge$ 12 \\
 & PFS6 & $\le$ 2, 4, 6 & $\le$ 1, 3, 5 & $\ge$ 12 & $\ge$ 19 \\
\bottomrule
\end{tabular}}
\tabnote{Interim entries correspond to information fractions 0.25, 0.50 and 0.75. Futility requires both criteria to cross their boundaries at the same look; either criterion suffices for a final claim. A dash denotes an unattainable futility boundary. Pooled claims are evaluated only after both cohorts reach their planned sizes; routing follows Section~\ref{sec2.3}.}
\end{table}

\newpage

\begin{center}\setlength{\tabcolsep}{2pt}
\fontsize{8pt}{9.6pt}\selectfont
\begin{longtable}{lcccccccccc}
\caption{Operating characteristics for co-primary endpoints.}\label{tab6}\addcontentsline{toc}{subsection}{Table S9: Operating characteristics for co-primary endpoints}\\
\toprule
\textbf{Sc.} & \textbf{Design} & \textbf{max $N$} & \textbf{$\mathrm{FWER}_{00}$} & \textbf{$P(S)$} & \textbf{$\max_{\mathrm{grid}}\mathrm{FWER}_{01}$} & \textbf{$\widehat{U}_S$} & \textbf{$\mathrm{POWER}_{01}$} & \textbf{$\mathrm{POWER}_{11}$} & \textbf{Disjunctive power} & \textbf{$\mathrm{ESS}_0$} \\
\midrule
\endfirsthead
\multicolumn{11}{l}{\small\textit{Table~\thetable{} (continued)}}\\
\toprule
\textbf{Sc.} & \textbf{Design} & \textbf{max $N$} & \textbf{$\mathrm{FWER}_{00}$} & \textbf{$P(S)$} & \textbf{$\max_{\mathrm{grid}}\mathrm{FWER}_{01}$} & \textbf{$\widehat{U}_S$} & \textbf{$\mathrm{POWER}_{01}$} & \textbf{$\mathrm{POWER}_{11}$} & \textbf{Disjunctive power} & \textbf{$\mathrm{ESS}_0$} \\
\midrule
\endhead
\endfoot
\bottomrule
\endlastfoot
S1 & BOP2-P & 120 & 0.0296 & 1.0000 & \textbf{1.0000} & 1.0000 & 0.553 & --- & 0.978 & 94.1 \\
 & BOP2-ENR-nE & 120 & 0.0520 & 1.0000 & \textbf{1.0000} & 1.0000 & 0.857 & --- & 0.962 & 81.6 \\
 & BOP2-ENR & 120 & 0.0330 & 0.0494 & \textbf{0.0497} & 0.0494 & 0.835 & 0.649 & 0.922 & 52.3 \\
S2 & BOP2-P & 40 & 0.0311 & 1.0000 & \textbf{0.9997} & 1.0000 & 0.673 & --- & 0.968 & 29.2 \\
 & BOP2-ENR-nE & 40 & 0.0419 & 1.0000 & \textbf{1.0000} & 1.0000 & 0.871 & --- & 0.957 & 28.2 \\
 & BOP2-ENR & 40 & 0.0276 & 0.0308 & \textbf{0.0315} & 0.0308 & 0.857 & 0.389 & 0.899 & 19.7 \\
S3 & BOP2-P & 60 & 0.0423 & 1.0000 & \textbf{0.9979} & 1.0000 & 0.591 & --- & 0.991 & 44.7 \\
 & BOP2-ENR-nE & 60 & 0.0535 & 1.0000 & \textbf{1.0000} & 1.0000 & 0.912 & --- & 0.988 & 45.0 \\
 & BOP2-ENR & 60 & 0.0314 & 0.0404 & \textbf{0.0410} & 0.0404 & 0.898 & 0.609 & 0.954 & 26.9 \\
S4 & BOP2-P & 60 & 0.0439 & 1.0000 & \textbf{0.9938} & 1.0000 & 0.616 & --- & 0.999 & 47.4 \\
 & BOP2-ENR-nE & 60 & 0.0592 & 1.0000 & \textbf{1.0000} & 1.0000 & 0.950 & --- & 0.997 & 47.7 \\
 & BOP2-ENR & 60 & \textbf{0.0369} & 0.0285 & 0.0290 & 0.0369 & 0.941 & 0.739 & 0.983 & 28.4 \\
S5 & BOP2-P & 60 & 0.0236 & 1.0000 & \textbf{1.0000} & 1.0000 & 0.694 & --- & 0.951 & 48.8 \\
 & BOP2-ENR-nE & 60 & 0.0653 & 1.0000 & \textbf{1.0000} & 1.0000 & 0.891 & --- & 0.976 & 47.9 \\
 & BOP2-ENR & 60 & \textbf{0.0444} & 0.0308 & 0.0309 & 0.0444 & 0.868 & 0.392 & 0.915 & 37.6 \\
S6 & BOP2-P & 80 & 0.0356 & 1.0000 & \textbf{1.0000} & 1.0000 & 0.820 & --- & 0.968 & 62.4 \\
 & BOP2-ENR-nE & 80 & 0.0329 & 1.0000 & \textbf{1.0000} & 1.0000 & 0.867 & --- & 0.961 & 63.3 \\
 & BOP2-ENR & 80 & 0.0242 & 0.0308 & \textbf{0.0315} & 0.0308 & 0.847 & 0.391 & 0.897 & 53.1 \\
S7 & BOP2-P & 40 & 0.0306 & 1.0000 & \textbf{0.9998} & 1.0000 & 0.562 & --- & 0.977 & 30.6 \\
 & BOP2-ENR-nE & 40 & 0.0607 & 1.0000 & \textbf{1.0000} & 1.0000 & 0.808 & --- & 0.990 & 40.0 \\
 & BOP2-ENR & 40 & 0.0316 & 0.0308 & \textbf{0.0318} & 0.0316 & 0.759 & 0.550 & 0.888 & 25.9 \\
S8 & BOP2-P & 40 & 0.0252 & 1.0000 & \textbf{0.9980} & 1.0000 & 0.575 & --- & 0.970 & 31.0 \\
 & BOP2-ENR-nE & 40 & 0.0431 & 1.0000 & \textbf{1.0000} & 1.0000 & 0.847 & --- & 0.975 & 31.8 \\
 & BOP2-ENR & 40 & 0.0337 & 0.0470 & \textbf{0.0460} & 0.0470 & 0.832 & 0.605 & 0.927 & 22.1 \\
\end{longtable}
\tabnote{Metrics and comparator interpretations follow Table~\ref{tab3}. BOP2-ENR-nE uses the subgroup quotas of BOP2-ENR; pooled BOP2 (BOP2-P) uses the same maximum total sample size and planned mixture weight, drawing all patients from the mixture (Section~\ref{sec3.2}). BOOST is not evaluated for this endpoint. $P(S)$ is exact; other probabilities use $10^5$ simulations. $\mathrm{POWER}_{11}$ is reported only for BOP2-ENR.}
\end{center}

\clearpage
\SIsec{Supporting tables: joint efficacy-toxicity monitoring}

\begin{table}[H]
\centering
\caption{Scenario parameters and subgroup sample sizes, joint efficacy-toxicity monitoring.}\label{tabSet}\addcontentsline{toc}{subsection}{Table S10: Scenario parameters and subgroup sample sizes, joint efficacy-toxicity monitoring}
{\fontsize{8.5pt}{10pt}\selectfont\setlength{\tabcolsep}{4pt}
\begin{tabular}{lccccccc}
\toprule
\textbf{Sc.} & \textbf{Null $(p_0^-,p_0^+)$} & \textbf{Alt. $(p_1^-,p_1^+)$} & \textbf{Tox. threshold $\phi_{0T}$} & \textbf{True tox. rate} & \textbf{$N^-$} & \textbf{$N^+$} & \textbf{Pooled null $\phi_0^{\mathrm{pool}}$} \\
\midrule
S1 & (0.03, 0.03) & (0.10, 0.15) & 0.30 & 0.10 & 80 & 40 & 0.030 \\
S2 & (0.05, 0.05) & (0.20, 0.25) & 0.30 & 0.10 & 20 & 20 & 0.050 \\
S3 & (0.10, 0.10) & (0.25, 0.35) & 0.30 & 0.10 & 40 & 20 & 0.100 \\
S4 & (0.20, 0.20) & (0.40, 0.50) & 0.30 & 0.10 & 40 & 20 & 0.200 \\
S5 & (0.05, 0.10) & (0.20, 0.25) & 0.30 & 0.10 & 20 & 40 & 0.083 \\
S6 & (0.05, 0.20) & (0.20, 0.35) & 0.30 & 0.10 & 20 & 60 & 0.163 \\
S7 & (0.05, 0.10) & (0.25, 0.30) & 0.30 & 0.10 & 20 & 20 & 0.075 \\
S8 & (0.10, 0.20) & (0.30, 0.45) & 0.30 & 0.10 & 20 & 20 & 0.150 \\
\bottomrule
\end{tabular}}
\tabnote{Response rates and sample sizes are those of Table~\ref{tab1}. The toxicity threshold $\phi_{0T} = 0.30$ and the tolerable (true) toxicity rate 0.10 apply in both subgroups in every scenario; on the toxicity-null face the rate is placed at 0.30 (Section~\ref{sec2.4}). Safety is never pooled, so only the response threshold is pooled.}
\end{table}

\begin{table}[H]
\centering
\caption{Sensitivity to the number of looks: efficacy-toxicity monitoring.}\label{tabS7}\addcontentsline{toc}{subsection}{Table S11: Sensitivity to the number of looks: efficacy-toxicity monitoring}
{\fontsize{9pt}{10.8pt}\selectfont
\begin{tabular}{lccccc}
\toprule
\shortstack{\textbf{Total}\\\textbf{looks $R$}} & \shortstack{\textbf{Interim}\\\textbf{futility looks}} & \shortstack{\textbf{Scenarios with profiled}\\\textbf{diagnostic $\mathrm{FWER}_S \le \alpha$}} & \shortstack{\textbf{Mean}\\\textbf{$\mathrm{POWER}_{01}$}} & \shortstack{\textbf{Mean}\\\textbf{disjunctive power}} & \shortstack{\textbf{Mean}\\\textbf{$\mathrm{ESS}_0$}} \\
\midrule
2 & 1 & 8/8 & 0.630 & 0.661 & 41.6 \\
3 & 2 & 8/8 & 0.630 & 0.674 & 38.0 \\
4 & 3 & 8/8 & 0.626 & 0.672 & 35.2 \\
5 & 4 & 8/8 & 0.623 & 0.666 & 34.0 \\
6 & 5 & 7/8 & 0.620 & 0.668 & 33.1 \\
\bottomrule
\end{tabular}}
\tabnote{Each scenario was recalibrated for each total number of looks $R$, including the final analysis. Power and sample-size entries are scenario averages. The count column summarizes the profiled diagnostic, not calibration feasibility (Appendix S2). Operating characteristics use $6 \times 10^4$ simulations per evaluated design.}
\end{table}

\begin{table}[H]
\centering
\caption{Calibrated parameters for efficacy-toxicity monitoring.}\label{tabS8}\addcontentsline{toc}{subsection}{Table S12: Calibrated parameters for efficacy-toxicity monitoring}
{\fontsize{9pt}{10.8pt}\selectfont
\begin{tabular}{lcccccc}
\toprule
\textbf{Sc.} & \textbf{NG $\lambda_-$} & \textbf{NG $\gamma_-$} & \textbf{PG $\lambda_+$} & \textbf{Efficacy $\lambda_E$} & \textbf{Safety $\lambda_T^-$} & \textbf{Safety $\lambda_T^+$} \\
\midrule
S1 & 0.920 & 0.35 & 0.05 & 0.900 & 0.970 & 0.970 \\
S2 & 0.950 & 0.35 & 0.02 & 0.940 & 0.950 & 0.950 \\
S3 & 0.950 & 0.35 & 0.35 & 0.880 & 0.970 & 0.950 \\
S4 & 0.985 & 0.35 & 0.02 & 0.970 & 0.970 & 0.950 \\
S5 & 0.950 & 0.35 & 0.02 & 0.950 & 0.950 & 0.970 \\
S6 & 0.950 & 0.35 & 0.20 & 0.950 & 0.950 & 0.970 \\
S7 & 0.950 & 0.35 & 0.05 & 0.950 & 0.950 & 0.950 \\
S8 & 0.950 & 0.35 & 0.05 & 0.940 & 0.950 & 0.950 \\
\bottomrule
\end{tabular}}
\tabnote{Futility cutoffs follow Table~\ref{tabS2}, with $\gamma_+ = 0.50$. $\lambda_E$ applies to efficacy; $\lambda_T^-$ and $\lambda_T^+$ are subgroup-specific final safety cutoffs, calibrated using exact binomial probabilities at toxicity rate 0.30 (Section~\ref{sec2.5b}).}
\end{table}

The calibration sits close to the binary case rather than the co-primary one: the disjunctive futility rule already stops the NG cohort readily, so $\gamma_-$ stays at 0.35 throughout and the gate needs no reshaping.

\begin{center}
\fontsize{9pt}{10.8pt}\selectfont
\begin{longtable}{lcccccc}
\caption{Enrichment timing and trial stopping: efficacy-toxicity monitoring.}\label{tabS9}\addcontentsline{toc}{subsection}{Table S13: Enrichment timing and trial stopping: efficacy-toxicity monitoring}\\
\toprule
\textbf{Sc.} & \textbf{Configuration} & \textbf{$L_1$} & \textbf{$L_2$} & \textbf{$L_3$} & \textbf{Never enriched} & \textbf{Trial stopped (PG futility)} \\
\midrule
\endfirsthead
\multicolumn{7}{l}{\small\textit{Table~\thetable{} (continued)}}\\
\toprule
\textbf{Sc.} & \textbf{Configuration} & \textbf{$L_1$} & \textbf{$L_2$} & \textbf{$L_3$} & \textbf{Never enriched} & \textbf{Trial stopped (PG futility)} \\
\midrule
\endhead
\endfoot
\bottomrule
\endlastfoot
S1 & global null & 88.0 & 2.5 & 1.1 & 8.5 & 54.7 \\
 & mixed & 88.0 & 5.2 & 2.3 & 4.5 & 3.9 \\
 & global alt & 40.0 & 3.2 & 1.2 & 55.6 & 3.9 \\
S2 & global null & 79.1 & 14.7 & 2.2 & 3.9 & 46.7 \\
 & mixed & 79.4 & 14.6 & 4.0 & 1.9 & 1.4 \\
 & global alt & 38.6 & 13.9 & 11.0 & 36.5 & 1.4 \\
S3 & global null & 30.6 & 6.0 & 1.5 & 61.9 & 71.0 \\
 & mixed & 65.7 & 13.7 & 4.6 & 16.0 & 13.0 \\
 & global alt & 26.1 & 7.9 & 3.8 & 62.3 & 13.0 \\
S4 & global null & 70.1 & 14.4 & 8.5 & 6.9 & 10.9 \\
 & mixed & 69.7 & 16.5 & 9.5 & 4.3 & 0.1 \\
 & global alt & 22.8 & 6.0 & 6.5 & 64.7 & 0.1 \\
S5 & global null & 79.1 & 12.9 & 3.8 & 4.2 & 12.4 \\
 & mixed & 79.1 & 14.7 & 4.2 & 1.9 & 0.3 \\
 & global alt & 38.5 & 14.0 & 11.2 & 36.3 & 0.3 \\
S6 & global null & 65.8 & 11.6 & 3.0 & 19.6 & 28.4 \\
 & mixed & 78.0 & 14.5 & 4.2 & 3.3 & 1.6 \\
 & global alt & 38.0 & 13.8 & 11.0 & 37.2 & 1.5 \\
S7 & global null & 79.3 & 9.4 & 2.8 & 8.5 & 35.1 \\
 & mixed & 79.3 & 14.1 & 4.1 & 2.5 & 2.9 \\
 & global alt & 30.1 & 10.4 & 10.3 & 49.2 & 2.9 \\
S8 & global null & 62.5 & 16.4 & 13.7 & 7.5 & 10.9 \\
 & mixed & 62.5 & 18.4 & 15.3 & 3.8 & 0.3 \\
 & global alt & 23.8 & 7.9 & 18.2 & 50.1 & 0.3 \\
\end{longtable}
\tabnote{Percentages of $10^5$ simulated trials; timing categories follow Table~\ref{tabS3}. PG stopping is a separate, overlapping measure, and ``Never enriched'' is not the certifying bound.}
\end{center}

\begin{table}[htbp]
\centering
\caption{Decision boundaries for efficacy-toxicity monitoring.}\label{tab8}\addcontentsline{toc}{subsection}{Table S14: Decision boundaries for efficacy-toxicity monitoring}
{\fontsize{9pt}{10.8pt}\selectfont
\begin{tabular}{lccccc}
\toprule
\textbf{Sc.} & \textbf{Criterion} & \textbf{Futility, NG} & \textbf{Futility, PG} & \textbf{PG-only claim} & \textbf{Pooled claim} \\
\midrule
S1 & ORR & $\le$ 1, 2, 3 & $\le$ ---, 0, 0 & $\ge$ 4 & $\ge$ 7 \\
 & Tox & $\ge$ 6, 11, 15 & $\ge$ 7, 11, 14 & $\le$ 6 & NG $\le$ 16; PG $\le$ 6 \\
S2 & ORR & $\le$ 0, 1, 2 & $\le$ ---, ---, 0 & $\ge$ 4 & $\ge$ 5 \\
 & Tox & $\ge$ 2, 3, 3 & $\ge$ 5, 7, 9 & $\le$ 2 & NG $\le$ 2; PG $\le$ 2 \\
S3 & ORR & $\le$ 1, 3, 5 & $\le$ 0, 0, 1 & $\ge$ 5 & $\ge$ 10 \\
 & Tox & $\ge$ 3, 5, 7 & $\ge$ 3, 5, 6 & $\le$ 2 & NG $\le$ 6; PG $\le$ 2 \\
S4 & ORR & $\le$ 2, 5, 9 & $\le$ ---, 0, 0 & $\ge$ 8 & $\ge$ 19 \\
 & Tox & $\ge$ 3, 5, 7 & $\ge$ 5, 7, 9 & $\le$ 2 & NG $\le$ 6; PG $\le$ 2 \\
S5 & ORR & $\le$ 0, 1, 2 & $\le$ ---, 0, 0 & $\ge$ 8 & $\ge$ 10 \\
 & Tox & $\ge$ 2, 3, 3 & $\ge$ 7, 11, 15 & $\le$ 6 & NG $\le$ 2; PG $\le$ 6 \\
S6 & ORR & $\le$ 0, 1, 2 & $\le$ 1, 3, 6 & $\ge$ 18 & $\ge$ 19 \\
 & Tox & $\ge$ 2, 3, 3 & $\ge$ 8, 12, 17 & $\le$ 11 & NG $\le$ 2; PG $\le$ 11 \\
S7 & ORR & $\le$ 0, 1, 2 & $\le$ ---, 0, 0 & $\ge$ 5 & $\ge$ 7 \\
 & Tox & $\ge$ 2, 3, 3 & $\ge$ 4, 6, 8 & $\le$ 2 & NG $\le$ 2; PG $\le$ 2 \\
S8 & ORR & $\le$ 0, 1, 3 & $\le$ ---, 0, 0 & $\ge$ 8 & $\ge$ 10 \\
 & Tox & $\ge$ 2, 3, 3 & $\ge$ 4, 6, 8 & $\le$ 2 & NG $\le$ 2; PG $\le$ 2 \\
\bottomrule
\end{tabular}}
\tabnote{Interim entries correspond to information fractions 0.25, 0.50 and 0.75. Futility occurs if response is at or below, or toxicity at or above, its boundary. A claim requires efficacy and safety. Pooled efficacy is evaluated only after both cohorts reach their planned sizes; safety must be established separately in each subgroup. A dash denotes an unattainable futility boundary; routing follows Section~\ref{sec2.3}.}
\end{table}

\newpage

\begin{center}\setlength{\tabcolsep}{2pt}
\fontsize{8pt}{9.6pt}\selectfont
\begin{longtable}{lcccccccccc}
\caption{Operating characteristics for efficacy-toxicity monitoring.}\label{tab9}\addcontentsline{toc}{subsection}{Table S15: Operating characteristics for efficacy-toxicity monitoring}\\
\toprule
\textbf{Sc.} & \textbf{Design} & \textbf{max $N$} & \textbf{$\mathrm{FWER}_{00}$} & \textbf{$P(S)$} & \textbf{$\max_{\mathrm{grid}}\mathrm{FWER}_{01}$} & \textbf{$\widehat{U}_S$} & \textbf{$\mathrm{POWER}_{01}$} & \textbf{$\mathrm{POWER}_{11}$} & \textbf{Disjunctive power} & \textbf{$\mathrm{ESS}_0$} \\
\midrule
\endfirsthead
\multicolumn{11}{l}{\small\textit{Table~\thetable{} (continued)}}\\
\toprule
\textbf{Sc.} & \textbf{Design} & \textbf{max $N$} & \textbf{$\mathrm{FWER}_{00}$} & \textbf{$P(S)$} & \textbf{$\max_{\mathrm{grid}}\mathrm{FWER}_{01}$} & \textbf{$\widehat{U}_S$} & \textbf{$\mathrm{POWER}_{01}$} & \textbf{$\mathrm{POWER}_{11}$} & \textbf{Disjunctive power} & \textbf{$\mathrm{ESS}_0$} \\
\midrule
\endhead
\endfoot
\bottomrule
\multicolumn{11}{@{}p{0.85\textwidth}@{}}{\footnotesize\raggedright\textit{Note:} {Metrics and comparator interpretations follow Table~\ref{tab3}. For BOP2-ENR, $P(S)$ is the exact maximum over the two NG-null face bounds, and $\mathrm{FWER}_{00}$ is maximized over admissible global-null corners. $\mathrm{ESS}_0$ and $\mathrm{POWER}_{01}$ use the efficacy-null/tolerable-toxicity configuration (toxicity rate 0.10). For BOP2-P, $\mathrm{FWER}_{00}$ is the larger of the any-claim probabilities on the two faces of its own pooled union null (efficacy-null face at toxicity 0; toxicity-null face at pooled efficacy 1), against which its single cutoff $\lambda_E$ is calibrated. Other probabilities use $10^5$ simulations; BOOST is not evaluated for this endpoint.}}\\
\endlastfoot
S1 & BOP2-P & 120 & 0.0415 & 1.0000 & \textbf{1.0000} & 1.0000 & 0.567 & --- & 0.952 & 68.5 \\
 & BOP2-ENR-nE & 120 & 0.0683 & 1.0000 & \textbf{1.0000} & 1.0000 & 0.789 & --- & 0.861 & 87.3 \\
 & BOP2-ENR & 120 & 0.0409 & 0.0426 & \textbf{0.0434} & 0.0426 & 0.776 & 0.481 & 0.824 & 52.5 \\
S2 & BOP2-P & 40 & 0.0222 & 1.0000 & \textbf{0.9882} & 1.0000 & 0.446 & --- & 0.759 & 18.1 \\
 & BOP2-ENR-nE & 40 & 0.0496 & 1.0000 & \textbf{0.9925} & 1.0000 & 0.542 & --- & 0.607 & 35.4 \\
 & BOP2-ENR & 40 & \textbf{0.0355} & 0.0340 & 0.0343 & 0.0355 & 0.526 & 0.205 & 0.567 & 24.1 \\
S3 & BOP2-P & 60 & 0.0283 & 1.0000 & \textbf{0.9917} & 1.0000 & 0.379 & --- & 0.895 & 32.1 \\
 & BOP2-ENR-nE & 60 & 0.0641 & 1.0000 & \textbf{0.9928} & 1.0000 & 0.552 & --- & 0.589 & 31.5 \\
 & BOP2-ENR & 60 & \textbf{0.0450} & 0.0450 & 0.0446 & 0.0450 & 0.546 & 0.321 & 0.575 & 22.0 \\
S4 & BOP2-P & 60 & 0.0283 & 1.0000 & \textbf{0.9815} & 1.0000 & 0.409 & --- & 0.947 & 35.7 \\
 & BOP2-ENR-nE & 60 & 0.0473 & 1.0000 & \textbf{0.9926} & 1.0000 & 0.591 & --- & 0.659 & 56.8 \\
 & BOP2-ENR & 60 & 0.0410 & 0.0451 & \textbf{0.0446} & 0.0451 & 0.588 & 0.414 & 0.642 & 33.5 \\
S5 & BOP2-P & 60 & 0.0283 & 1.0000 & \textbf{0.9999} & 1.0000 & 0.651 & --- & 0.890 & 36.1 \\
 & BOP2-ENR-nE & 60 & 0.0497 & 1.0000 & \textbf{1.0000} & 1.0000 & 0.740 & --- & 0.805 & 56.4 \\
 & BOP2-ENR & 60 & \textbf{0.0460} & 0.0340 & 0.0340 & 0.0460 & 0.737 & 0.266 & 0.780 & 44.0 \\
S6 & BOP2-P & 80 & 0.0288 & 1.0000 & \textbf{1.0000} & 1.0000 & 0.703 & --- & 0.887 & 47.4 \\
 & BOP2-ENR-nE & 80 & 0.0567 & 1.0000 & \textbf{1.0000} & 1.0000 & 0.824 & --- & 0.888 & 66.8 \\
 & BOP2-ENR & 80 & \textbf{0.0463} & 0.0340 & 0.0340 & 0.0463 & 0.814 & 0.288 & 0.854 & 56.2 \\
S7 & BOP2-P & 40 & 0.0250 & 1.0000 & \textbf{0.9882} & 1.0000 & 0.467 & --- & 0.819 & 21.7 \\
 & BOP2-ENR-nE & 40 & 0.0533 & 1.0000 & \textbf{0.9928} & 1.0000 & 0.518 & --- & 0.593 & 33.0 \\
 & BOP2-ENR & 40 & \textbf{0.0477} & 0.0340 & 0.0334 & 0.0477 & 0.514 & 0.265 & 0.568 & 22.8 \\
S8 & BOP2-P & 40 & 0.0259 & 1.0000 & \textbf{0.9832} & 1.0000 & 0.455 & --- & 0.791 & 21.4 \\
 & BOP2-ENR-nE & 40 & 0.0756 & 1.0000 & \textbf{0.9928} & 1.0000 & 0.531 & --- & 0.608 & 37.9 \\
 & BOP2-ENR & 40 & \textbf{0.0472} & 0.0463 & 0.0461 & 0.0472 & 0.512 & 0.279 & 0.576 & 26.8 \\
\end{longtable}
\end{center}

\clearpage
\SIsec{Supporting tables: strong control, verification and sensitivity}

\begin{table}[H]
\centering
\caption{False PG claims in the excluded reverse-mixed configuration.}\label{tabS10}\addcontentsline{toc}{subsection}{Table S16: False PG claims in the excluded reverse-mixed configuration}
{\fontsize{9pt}{10.8pt}\selectfont
\begin{tabular}{lccc}
\toprule
\textbf{Scenario} & \textbf{$(p^-, p^+)$} & \textbf{P(any claim)} & \textbf{P(joint claim)} \\
\midrule
S5 & (0.10, 0.10) & 0.073 & 0.039 \\
S6 & (0.20, 0.20) & 0.161 & 0.138 \\
S7 & (0.10, 0.10) & 0.070 & 0.040 \\
S8 & (0.20, 0.20) & 0.128 & 0.105 \\
\bottomrule
\end{tabular}}
\tabnote{Binary endpoint, $p^- = p^+ = \phi_0^+$: NG is active and PG is null. This configuration is excluded by the monotone-activity assumption. Every claim includes PG, so the any-claim probability is the false PG-claim rate; $4 \times 10^5$ simulations.}
\end{table}

\begin{table}[H]
\centering
\caption{Locally re-optimized BOOST designs.}\label{tabS11}\addcontentsline{toc}{subsection}{Table S17: Locally re-optimized BOOST designs}
{\fontsize{9pt}{10.8pt}\selectfont
\begin{tabular}{lccccc}
\toprule
\textbf{Sc.} & \textbf{Variant} & \textbf{$(r_1^-, r_1^+, r_2^{\mathrm{pool}}, r_2^+)$} & \textbf{FWER (exact)} & \textbf{Power} & \textbf{$\mathrm{ESS}_0$} \\
\midrule
S1 & weak & (1, 1, 6, 3) & 0.043 & 0.820 & 48.3 \\
S1 & strong & (6, 0, 3, 3) & 0.024 & 0.829 & 58.0 \\
S2 & weak & (0, 1, 4, 2) & 0.044 & 0.840 & 19.8 \\
S2 & strong & (2, 0, 3, 3) & 0.044 & 0.848 & 26.8 \\
S3 & weak & (2, 1, 9, 4) & 0.043 & 0.804 & 25.5 \\
S3 & strong & (5, 0, 6, 4) & 0.028 & 0.814 & 30.2 \\
S4 & weak & (4, 1, 14, 5) & 0.049 & 0.802 & 18.6 \\
S4 & strong & (8, 1, 13, 7) & 0.023 & 0.821 & 26.6 \\
S5 & weak & (0, 3, 8, 7) & 0.039 & 0.812 & 28.8 \\
S5 & strong & (2, 1, 5, 7) & 0.040 & 0.804 & 37.8 \\
S6 & weak & (3, 4, 17, 19) & 0.037 & 0.820 & 45.9 \\
S6 & strong & (6, 5, 16, 18) & 0.044 & 0.830 & 48.7 \\
S7 & weak & (1, 1, 5, 5) & 0.042 & 0.823 & 19.1 \\
S8 & weak & (2, 2, 10, 6) & 0.050 & 0.808 & 17.3 \\
S8 & strong & (3, 3, 9, 8) & 0.031 & 0.802 & 21.4 \\
\bottomrule
\end{tabular}}
\tabnote{Sample sizes remain at published values; boundaries were searched within $\pm 3$ of the published values to minimize $\mathrm{ESS}_0$, subject to disjunctive power $\ge 0.80$ at the design alternative and the specified error constraint, ties being resolved by scan order. FWER denotes $\mathrm{FWER}_{00}$ for weak-control rows and $\max\{\mathrm{FWER}_{00}, \mathrm{FWER}_{01}(0.99)\}$ for strong-control rows, the mixed configuration being evaluated at $\psi = 0.99$. Probabilities are exact. These are local solutions; weak-control rows do not constrain mixed errors. Boundary conventions and the omitted S7 strong-control design are described in Appendix S3.}
\end{table}

\begin{table}[H]
\centering
\caption{Exact error bounds and independent Monte Carlo verification.}\label{tabS12}\addcontentsline{toc}{subsection}{Table S18: Exact error bounds and independent Monte Carlo verification}
{\fontsize{8pt}{9.6pt}\selectfont
\begin{tabular}{llcccccccccc}
\toprule
& & & \multicolumn{2}{c}{\textbf{faces (eff-tox)}} & \multicolumn{3}{c}{\textbf{$\max_{\mathrm{grid}}\mathrm{FWER}_{01}$}} & \multicolumn{2}{c}{\textbf{$\mathrm{FWER}_{00}$}} & \\
\cmidrule(lr){4-5}\cmidrule(lr){6-8}\cmidrule(lr){9-10}
\textbf{Sc.} & \textbf{Endpoint} & \textbf{bound} & \textbf{efficacy} & \textbf{toxicity} & \textbf{reported} & \textbf{seed 99} & \textbf{95\% UCL} & \textbf{seed 99} & \textbf{95\% UCL} & \textbf{$U_S^{\mathrm{UCL}}$} \\
\midrule
S1 & Binary & 0.0426 & --- & --- & 0.0432 & 0.0430 & 0.0437 & 0.0410 & 0.0415 & 0.0426 \\
S2 & Binary & 0.0237 & --- & --- & 0.0239 & 0.0241 & 0.0246 & 0.0239 & 0.0243 & 0.0243 \\
S3 & Binary & 0.0450 & --- & --- & 0.0456 & 0.0454 & 0.0461 & 0.0444 & 0.0450 & 0.0450 \\
S4 & Binary & 0.0451 & --- & --- & 0.0458 & 0.0451 & 0.0458 & 0.0389 & 0.0394 & 0.0451 \\
S5 & Binary & 0.0237 & --- & --- & 0.0239 & 0.0241 & 0.0246 & 0.0442 & 0.0448 & 0.0448 \\
S6 & Binary & 0.0237 & --- & --- & 0.0239 & 0.0241 & 0.0246 & 0.0449 & 0.0454 & 0.0454 \\
S7 & Binary & 0.0237 & --- & --- & 0.0239 & 0.0240 & 0.0245 & 0.0412 & 0.0417 & 0.0417 \\
S8 & Binary & 0.0463 & --- & --- & 0.0466 & 0.0467 & 0.0474 & 0.0398 & 0.0403 & 0.0463 \\
S1 & Co-primary & 0.0494 & --- & --- & 0.0497 & 0.0494 & 0.0501 & 0.0332 & 0.0337 & 0.0494 \\
S2 & Co-primary & 0.0308 & --- & --- & 0.0315 & 0.0307 & 0.0313 & 0.0278 & 0.0282 & 0.0308 \\
S3 & Co-primary & 0.0404 & --- & --- & 0.0410 & 0.0405 & 0.0412 & 0.0320 & 0.0324 & 0.0404 \\
S4 & Co-primary & 0.0285 & --- & --- & 0.0290 & 0.0286 & 0.0291 & 0.0364 & 0.0369 & 0.0369 \\
S5 & Co-primary & 0.0308 & --- & --- & 0.0309 & 0.0312 & 0.0318 & 0.0448 & 0.0453 & 0.0453 \\
S6 & Co-primary & 0.0308 & --- & --- & 0.0315 & 0.0312 & 0.0318 & 0.0240 & 0.0244 & 0.0308 \\
S7 & Co-primary & 0.0308 & --- & --- & 0.0318 & 0.0307 & 0.0313 & 0.0317 & 0.0322 & 0.0322 \\
S8 & Co-primary & 0.0470 & --- & --- & 0.0460 & 0.0472 & 0.0480 & 0.0337 & 0.0342 & 0.0470 \\
S1 & Efficacy-toxicity & 0.0426 & 0.0426 & 0.0253 & 0.0434 & 0.0428 & 0.0436 & 0.0416 & 0.0423 & 0.0426 \\
S2 & Efficacy-toxicity & 0.0340 & 0.0237 & 0.0340 & 0.0343 & 0.0336 & 0.0343 & 0.0356 & 0.0363 & 0.0363 \\
S3 & Efficacy-toxicity & 0.0450 & 0.0450 & 0.0201 & 0.0446 & 0.0444 & 0.0452 & 0.0445 & 0.0453 & 0.0453 \\
S4 & Efficacy-toxicity & 0.0451 & 0.0451 & 0.0201 & 0.0446 & 0.0454 & 0.0462 & 0.0412 & 0.0419 & 0.0451 \\
S5 & Efficacy-toxicity & 0.0340 & 0.0237 & 0.0340 & 0.0340 & 0.0340 & 0.0347 & 0.0449 & 0.0457 & 0.0457 \\
S6 & Efficacy-toxicity & 0.0340 & 0.0237 & 0.0340 & 0.0340 & 0.0344 & 0.0351 & 0.0463 & 0.0471 & 0.0471 \\
S7 & Efficacy-toxicity & 0.0340 & 0.0237 & 0.0340 & 0.0334 & 0.0332 & 0.0339 & 0.0470 & 0.0478 & 0.0478 \\
S8 & Efficacy-toxicity & 0.0463 & 0.0463 & 0.0340 & 0.0461 & 0.0465 & 0.0473 & 0.0472 & 0.0479 & 0.0479 \\
\bottomrule
\end{tabular}}
\tabnote{The reported bound is exact; for efficacy-toxicity it is the larger of the efficacy-face and toxicity-face values shown. Original estimates use $10^5$ simulations (seed 77); verification uses $4 \times 10^5$ (seed 99). UCLs are one-sided 95\% Clopper--Pearson limits, Bonferroni-adjusted within each design over null corners for $\mathrm{FWER}_{00}$ and over faces and grid points for $\max_{\mathrm{grid}}\mathrm{FWER}_{01}$. $U_S^{\mathrm{UCL}} = \max\{\mathrm{UCL}_{00}, P(S)\}$ combines the global-null limit with the exact mixed-error bound. The profiled diagnostic and its UCL do not replace that bound: its adjusted UCL exceeds 0.05 in one case, co-primary S1 (0.0501, exact bound 0.0494), where point estimates exceed the bound by at most 0.0005, within Monte Carlo error. Joint verification across designs is described in Section~\ref{sec3.3}.}
\end{table}

\begin{table}[H]
\centering
\caption{Sensitivity to the efficacy cutoff in co-primary scenario S3.}\label{tabS13}\addcontentsline{toc}{subsection}{Table S19: Sensitivity to the efficacy cutoff in co-primary scenario S3}
{\fontsize{9pt}{10.8pt}\selectfont
\begin{tabular}{cccc}
\toprule
\textbf{$\lambda_E$} & \textbf{$\mathrm{FWER}_{00}$} & \textbf{$\max_{\mathrm{grid}}\mathrm{FWER}_{01}$} & \textbf{$P(S)$} \\
\midrule
0.96 & 0.0860 & 0.0410 & 0.0404 \\
0.97 & 0.0314 & 0.0410 & 0.0404 \\
0.98 & 0.0297 & 0.0410 & 0.0404 \\
\bottomrule
\end{tabular}}
\tabnote{Futility parameters are fixed at the Table~\ref{tabS5} values. Error rates use $10^5$ simulations; $P(S)$ is exact.}
\end{table}

\begin{table}[H]
\centering
\caption{Exact ceilings and observed joint-claim power.}\label{tabS14}\addcontentsline{toc}{subsection}{Table S20: Exact ceilings and observed joint-claim power}
{\fontsize{9pt}{10.8pt}\selectfont
\begin{tabular}{llccccc}
\toprule
\textbf{Sc.} & \textbf{Endpoint} & \textbf{$P(S^{\ast})$ exact} & \textbf{ceiling with safety} & \textbf{$\mathrm{POWER}_{11}$ reported} & \textbf{seed 99} & \textbf{Difference} \\
\midrule
S1 & Binary & 0.561 & --- & 0.540 & 0.540 & -0.022 \\
S2 & Binary & 0.461 & --- & 0.454 & 0.454 & -0.007 \\
S3 & Binary & 0.631 & --- & 0.555 & 0.555 & -0.076 \\
S4 & Binary & 0.720 & --- & 0.687 & 0.687 & -0.033 \\
S5 & Binary & 0.461 & --- & 0.431 & 0.432 & -0.030 \\
S6 & Binary & 0.461 & --- & 0.443 & 0.443 & -0.018 \\
S7 & Binary & 0.617 & --- & 0.514 & 0.513 & -0.104 \\
S8 & Binary & 0.634 & --- & 0.620 & 0.621 & -0.014 \\
S1 & Co-primary & 0.672 & --- & 0.649 & 0.649 & -0.023 \\
S2 & Co-primary & 0.401 & --- & 0.389 & 0.391 & -0.011 \\
S3 & Co-primary & 0.613 & --- & 0.609 & 0.609 & -0.004 \\
S4 & Co-primary & 0.740 & --- & 0.739 & 0.739 & -0.001 \\
S5 & Co-primary & 0.401 & --- & 0.392 & 0.396 & -0.005 \\
S6 & Co-primary & 0.401 & --- & 0.391 & 0.392 & -0.009 \\
S7 & Co-primary & 0.550 & --- & 0.550 & 0.551 & +0.000 \\
S8 & Co-primary & 0.613 & --- & 0.605 & 0.606 & -0.008 \\
S1 & Efficacy-toxicity & 0.555 & 0.499 & 0.481 & 0.480 & -0.019 \\
S2 & Efficacy-toxicity & 0.365 & 0.207 & 0.205 & 0.204 & -0.002 \\
S3 & Efficacy-toxicity & 0.570 & 0.364 & 0.321 & 0.321 & -0.043 \\
S4 & Efficacy-toxicity & 0.650 & 0.415 & 0.414 & 0.414 & -0.001 \\
S5 & Efficacy-toxicity & 0.365 & 0.275 & 0.266 & 0.267 & -0.008 \\
S6 & Efficacy-toxicity & 0.365 & 0.301 & 0.288 & 0.291 & -0.010 \\
S7 & Efficacy-toxicity & 0.488 & 0.276 & 0.265 & 0.267 & -0.010 \\
S8 & Efficacy-toxicity & 0.501 & 0.284 & 0.279 & 0.282 & -0.002 \\
\bottomrule
\end{tabular}}
\tabnote{$P(S^{\ast})$ is exact NG survival under its alternative. For efficacy-toxicity, the refined ceiling is $P(S^{\ast} \cap \text{NG safe})\,P(\text{PG safe})$. Reported power is from Tables~\ref{tab3}, \ref{tab6} and \ref{tab9}; verification uses $4 \times 10^5$ simulations (seed 99). The difference column compares verified power with the applicable ceiling, calculated before rounding; the gap includes PG futility stops as well as pooled efficacy claims not reached.}
\end{table}

\clearpage
\begin{table}[H]
\centering
\caption{Error-control sensitivity to endpoint association.}\label{tabS15}\addcontentsline{toc}{subsection}{Table S21: Error-control sensitivity to endpoint association}
{\fontsize{7.5pt}{9pt}\selectfont
\setlength{\tabcolsep}{1.6pt}
\begin{tabular}{ll|ccccc|ccccc|ccccc|c}
\toprule
& & \multicolumn{5}{c|}{\textbf{exact bound, OR $=$}} & \multicolumn{5}{c|}{\textbf{toxicity face at $(p_1^-, \phi_{0T})$, OR $=$}} & \multicolumn{5}{c|}{\textbf{$\mathrm{FWER}_{00}$ (corner maximum), OR $=$}} & \textbf{max} \\
\textbf{Sc.} & \textbf{Endpoint} & 0.25 & 0.5 & 1 & 2 & 4 & 0.25 & 0.5 & 1 & 2 & 4 & 0.25 & 0.5 & 1 & 2 & 4 & \textbf{UCL} \\
\midrule
S1 & Co-prim. & 0.0496 & 0.0495 & 0.0494 & 0.0491 & 0.0487 & --- & --- & --- & --- & --- & 0.0343 & 0.0339 & 0.0334 & 0.0334 & 0.0327 & 0.0353 \\
S2 & Co-prim. & 0.0309 & 0.0309 & 0.0308 & 0.0307 & 0.0303 & --- & --- & --- & --- & --- & 0.0282 & 0.0284 & 0.0280 & 0.0277 & 0.0268 & 0.0293 \\
S3 & Co-prim. & 0.0403 & 0.0404 & 0.0404 & 0.0402 & 0.0397 & --- & --- & --- & --- & --- & 0.0320 & 0.0319 & 0.0316 & 0.0318 & 0.0312 & 0.0329 \\
S4 & Co-prim. & 0.0281 & 0.0283 & 0.0285 & 0.0283 & 0.0279 & --- & --- & --- & --- & --- & 0.0362 & 0.0367 & 0.0355 & 0.0363 & 0.0351 & 0.0377 \\
S5 & Co-prim. & 0.0309 & 0.0309 & 0.0308 & 0.0307 & 0.0303 & --- & --- & --- & --- & --- & 0.0458 & 0.0459 & 0.0451 & 0.0442 & 0.0439 & 0.0470 \\
S6 & Co-prim. & 0.0309 & 0.0309 & 0.0308 & 0.0307 & 0.0303 & --- & --- & --- & --- & --- & 0.0246 & 0.0247 & 0.0245 & 0.0241 & 0.0234 & 0.0255 \\
S7 & Co-prim. & 0.0309 & 0.0309 & 0.0308 & 0.0307 & 0.0303 & --- & --- & --- & --- & --- & 0.0319 & 0.0315 & 0.0318 & 0.0322 & 0.0315 & 0.0331 \\
S8 & Co-prim. & 0.0472 & 0.0472 & 0.0470 & 0.0465 & 0.0455 & --- & --- & --- & --- & --- & 0.0344 & 0.0333 & 0.0343 & 0.0340 & 0.0332 & 0.0354 \\
S1 & Eff-tox & 0.0426 & 0.0426 & 0.0426 & 0.0426 & 0.0426 & 0.0162 & 0.0154 & 0.0142 & 0.0126 & 0.0108 & 0.0414 & 0.0414 & 0.0414 & 0.0414 & 0.0414 & 0.0429 \\
S2 & Eff-tox & 0.0340 & 0.0340 & 0.0340 & 0.0340 & 0.0340 & 0.0198 & 0.0181 & 0.0157 & 0.0127 & 0.0095 & 0.0348 & 0.0348 & 0.0348 & 0.0348 & 0.0348 & 0.0361 \\
S3 & Eff-tox & 0.0450 & 0.0450 & 0.0450 & 0.0450 & 0.0450 & 0.0153 & 0.0142 & 0.0127 & 0.0107 & 0.0085 & 0.0444 & 0.0444 & 0.0444 & 0.0444 & 0.0444 & 0.0458 \\
S4 & Eff-tox & 0.0451 & 0.0451 & 0.0451 & 0.0451 & 0.0451 & 0.0173 & 0.0161 & 0.0145 & 0.0124 & 0.0101 & 0.0411 & 0.0411 & 0.0411 & 0.0411 & 0.0411 & 0.0425 \\
S5 & Eff-tox & 0.0340 & 0.0340 & 0.0340 & 0.0340 & 0.0340 & 0.0198 & 0.0181 & 0.0157 & 0.0127 & 0.0095 & 0.0449 & 0.0449 & 0.0449 & 0.0449 & 0.0449 & 0.0464 \\
S6 & Eff-tox & 0.0340 & 0.0340 & 0.0340 & 0.0340 & 0.0340 & 0.0198 & 0.0181 & 0.0157 & 0.0127 & 0.0095 & 0.0493 & 0.0474 & 0.0466 & 0.0449 & 0.0449 & 0.0509 \\
S7 & Eff-tox & 0.0340 & 0.0340 & 0.0340 & 0.0340 & 0.0340 & 0.0250 & 0.0233 & 0.0210 & 0.0180 & 0.0145 & 0.0471 & 0.0471 & 0.0471 & 0.0471 & 0.0471 & 0.0486 \\
S8 & Eff-tox & 0.0463 & 0.0463 & 0.0463 & 0.0463 & 0.0463 & 0.0262 & 0.0243 & 0.0216 & 0.0182 & 0.0144 & 0.0472 & 0.0472 & 0.0472 & 0.0472 & 0.0472 & 0.0487 \\
\bottomrule
\end{tabular}}
\tabnote{Designs calibrated under independence are evaluated at fixed marginals with association varied through a Plackett odds ratio (OR), common to both subgroups and all null corners. Bounds are exact; $\mathrm{FWER}_{00}$ uses $10^5$ simulations per admissible corner. The additional toxicity-face bound is evaluated at $(p_1^-, \phi_{0T})$. The final column is the maximum of the per-OR, corner-adjusted one-sided 95\% Clopper--Pearson limits; it is not simultaneous across OR values.}
\end{table}

\begin{table}[H]
\centering
\caption{Power and sample-size sensitivity to endpoint association.}\label{tabS16}\addcontentsline{toc}{subsection}{Table S22: Power and sample-size sensitivity to endpoint association}
{\fontsize{8pt}{9.6pt}\selectfont
\begin{tabular}{llccccc|ccccc}
\toprule
& & \multicolumn{5}{c|}{\textbf{$\mathrm{POWER}_{01}$, OR $=$}} & \multicolumn{5}{c}{\textbf{$\mathrm{ESS}_0$, OR $=$}} \\
\textbf{Sc.} & \textbf{Endpoint} & 0.25 & 0.5 & 1 & 2 & 4 & 0.25 & 0.5 & 1 & 2 & 4 \\
\midrule
S1 & Co-prim. & 0.861 & 0.851 & 0.836 & 0.816 & 0.797 & 52.9 & 52.7 & 52.4 & 51.9 & 51.4 \\
S2 & Co-prim. & 0.894 & 0.878 & 0.857 & 0.838 & 0.815 & 19.9 & 19.8 & 19.7 & 19.5 & 19.2 \\
S3 & Co-prim. & 0.939 & 0.919 & 0.899 & 0.878 & 0.855 & 27.3 & 27.1 & 26.9 & 26.6 & 26.2 \\
S4 & Co-prim. & 0.971 & 0.958 & 0.941 & 0.922 & 0.906 & 29.2 & 28.9 & 28.5 & 28.0 & 27.4 \\
S5 & Co-prim. & 0.903 & 0.887 & 0.869 & 0.848 & 0.825 & 38.2 & 38.0 & 37.6 & 37.1 & 36.6 \\
S6 & Co-prim. & 0.894 & 0.871 & 0.844 & 0.822 & 0.797 & 54.9 & 54.1 & 53.1 & 52.0 & 50.9 \\
S7 & Co-prim. & 0.805 & 0.782 & 0.761 & 0.735 & 0.713 & 25.9 & 25.9 & 25.9 & 25.8 & 25.8 \\
S8 & Co-prim. & 0.887 & 0.861 & 0.836 & 0.809 & 0.785 & 22.8 & 22.5 & 22.1 & 21.7 & 21.3 \\
S1 & Eff-tox & 0.780 & 0.778 & 0.777 & 0.774 & 0.770 & 52.5 & 52.5 & 52.4 & 52.4 & 52.5 \\
S2 & Eff-tox & 0.539 & 0.533 & 0.528 & 0.515 & 0.499 & 24.1 & 24.1 & 24.1 & 24.0 & 24.0 \\
S3 & Eff-tox & 0.562 & 0.557 & 0.548 & 0.540 & 0.531 & 22.1 & 22.0 & 22.0 & 22.0 & 21.9 \\
S4 & Eff-tox & 0.606 & 0.598 & 0.590 & 0.581 & 0.575 & 33.7 & 33.6 & 33.5 & 33.4 & 33.3 \\
S5 & Eff-tox & 0.743 & 0.744 & 0.740 & 0.731 & 0.726 & 44.0 & 43.9 & 43.9 & 43.9 & 43.9 \\
S6 & Eff-tox & 0.813 & 0.813 & 0.813 & 0.813 & 0.812 & 56.3 & 56.3 & 56.2 & 56.2 & 56.1 \\
S7 & Eff-tox & 0.531 & 0.525 & 0.514 & 0.501 & 0.489 & 22.8 & 22.8 & 22.8 & 22.8 & 22.7 \\
S8 & Eff-tox & 0.535 & 0.525 & 0.510 & 0.502 & 0.491 & 26.9 & 26.9 & 26.8 & 26.8 & 26.7 \\
\bottomrule
\end{tabular}}
\tabnote{Association settings follow Table~\ref{tabS15}. Entries are $\mathrm{POWER}_{01}$ and $\mathrm{ESS}_0$; efficacy-toxicity uses the efficacy-null/tolerable-toxicity configuration.}
\end{table}

\end{document}